\documentclass[aps,pre,twocolumn]{revtex4}

\usepackage{amsmath}
\usepackage{bm}
\usepackage{graphicx}

\DeclareMathOperator{\sym}{sym}
\DeclareMathOperator{\grad}{grad}

\DeclareMathOperator{\divergence}{div}

\newcommand{\Div}{\divergence}

\newcommand{\mb}[1]{\mathbf{#1}}   
\newcommand{\Ev}[2]{\left.{#1}\right|_{#2}}
\newcommand{\Tp}[1]{{}^{\mathrm{t}}{#1}}

\newcommand{\rD}{\partial}
\newcommand{\dt}{\partial_t}
\newcommand{\dd}[2]{\frac{\partial #1}{\partial #2}}

\newcommand{\D}{\mathrm{d}}
\newcommand{\DD}[2]{\frac{\D #1}{\D #2}}  

\newcommand{\Eps}{\varepsilon}
\newcommand{\ld}{\lambda}

\newcommand{\ze}{\zeta}

\newcommand{\xS}{x_\text{S}}
\newcommand{\xD}{x_\text{D}}
\newcommand{\Nat}{\natural}
\newcommand{\xNat}{x^\Nat}
\newcommand{\DrNat}{\D\rvect^\Nat}

\newcommand{\rvect}{\mb{r}}
\newcommand{\BoldXi}{\bm{\xi}}
\newcommand{\TensorP}{\mathsf{P}}
\newcommand{\TensorSigma}{\tensor{\sigma}}
\newcommand{\detG}{\det{\mathsf{g}}}
\newcommand{\detGNat}{\det{\mathsf{g}^\Nat}}

\newcommand{\nd}{{n_{\text{d}}}}

\newcommand{\B}[1]{{\rD_{#1}}\rvect}
\newcommand{\Dx}{{\rD_\xi}}

\newcommand{\Dz}{{\rD_\ze}}
\newcommand{\Di}{{\rD_i}}
\newcommand{\Dj}{{\rD_j}}
\newcommand{\Dk}{{\rD_k}}
\newcommand{\dXi}{\D\xi}
\newcommand{\Cartesian}{\text{C}}

\newcommand{\Yield}{\sigma_{\text{Y}}}
\newcommand{\zYield}{\ze_{\text{Y}}}
\newcommand{\etaP}{\eta_{\text{p}}}
\newcommand{\Ell}{\ell_0}
\newcommand{\Tau}{\tau_0}

\newcommand{\patm}{p_{\text{atm}}}
\newcommand{\rhoGxH}{\rho G_x H}

\newcommand{\zm}{z_{\text{m}}}

\newcommand{\muS}{\mu_{\text{s}}}
\newcommand{\KernelG}[1]{\mathcal{G}(#1)}
\newcommand{\Temperature}{\mathcal{T}}

\begin{document}

\title{Continuum theory of memory effect
       in crack patterns of drying pastes}
\author{\surname{Ooshida} Takeshi}  
\email[E-mail:~]{ooshida@damp.tottori-u.ac.jp}
\affiliation{%
  Department of Applied Mathematics and Physics,
  Tottori University, JP-680-8552, Japan
}
\date{\today}

\begin{abstract}
 A possible clarification
 of memory effect observed in crack patterns of drying pastes
 [A.~Nakahara and Y.~Matsuo,
 J.~Phys.~Soc.~Japan \textbf{74}, 1362 (2005)]
 is presented 
 in terms of a macroscopic elastoplastic model of isotropic pastes.
 We study flows driven by steady gravitational force
 instead of external oscillation.
 The model predicts creation of residual tension
 in favor of cracks perpendicular to the flow direction,
 thus causing the same type of memory effect 
 as that reported by Nakahara and Matsuo
 for oscillated $\mathrm{CaCO_3}$ pastes.
\end{abstract}

\maketitle

\section{Introduction}
\label{sec:intro}

As the plastic behavior of soft glassy materials has been
  attracting increasing interest~\cite{Carmen-Miguel.Book2006},
  it was reported by Nakahara and 
  Matsuo~\cite{Nakahara.Bussei81,Nakahara.JPSJ74,Nakahara.JStat2006}
  that a drying paste exhibits a memory effect.
They observed a drying process for a paste
  containing calcium carbonate ($\mathrm{CaCO_3}$) and water
  in a shallow container
  in order to study the resulting crack pattern.
The crack pattern was typically found to be isotropic,
  but they discovered a way to introduce anisotropy into the paste
  \emph{before} the drying process commences:
  by applying a horizontal oscillation to the container
  immediately after the paste is poured into it,
  a memory of the oscillation is imprinted into the paste,
  which determines how it should break in the future.

Through systematic experiments,
  Nakahara and Matsuo also found
  that plasticity is essential to the memory effect
  in $\mathrm{CaCO_3}$ pastes.
No memory effect is observed
  if the strength of the applied oscillation
  is below the threshold value
  corresponding to the plastic yield stress of the paste.
Just above the threshold value,
  the paste remembers the oscillation that caused the plastic flow,
  developing cracks perpendicular to the direction of the oscillation.
If the oscillation is too strong or the paste contains too much water,
  waves and global flows are induced,
  eliminating the memory effect.
A different kind of paste (mixture
  of magnesium carbonate hydroxide with water)~\cite{Nakahara.PRE74}
  exhibits
  not only a memory effect similar to that of $\mathrm{CaCO_3}$
  that occurs just above the threshold of plastic flows
  and causes cracks perpendicular to the external oscillation,
  but also a different type of memory effect in its water-rich condition
  where the cracks are parallel to the direction
  of the global laminar flow caused by the oscillation.
Too strong an oscillation and too much water
  also destroy the memory in this paste,
  with the emergence of chaotic, turbulence-like flows~\footnote{%
    According to Nakahara (private communication),
    a typical value of the Reynolds number in such cases is 
    $R_H = 100$ on the basis of the layer thickness $H$,
    or $R_L = 2000$
    on the basis of the horizontal length scale $L$ of the container.
    While $R_H = 100$ is typically not large enough
    to cause a transition to turbulence (in the usual sense of the word),
    we may expect a different kind of ``turbulence''
    such as a two-dimensional chaotic flow
    maintained by horizontal forcing. 
  }
  characterized by fluid motion in every direction.

Here we focus our attention
  on the former type of memory effect
  that causes cracks perpendicular to the external forcing,
  which we refer to as the Type-I Nakahara effect.
The latter type, which could be called the Type-II Nakahara effect,
  will be discussed only briefly.

Although it is certain
  that the memory effect in $\mathrm{CaCO_3}$ pastes
  originates from plastic flow,
  it is unclear which aspects of the plastic flow are essential.
More specifically,
  because the role played by the unsteadiness of the flow
  is not fully understood,
  it is unknown 
  whether a slope flow, in which the external forcing is steady,
  can cause a memory effect.
To answer this question,
  coworkers of the present author have started
  an experimental study on the slope flow of $\mathrm{CaCO_3}$ paste.
Paste supplied through a funnel
  is driven downstream by gravitational force,
  and when the supply is stopped,
  the paste ``freezes'' at some finite thickness
  due to the finite yield stress.
Preliminary results suggest 
  the presence of a memory effect (Type-I Nakahara effect),
  where the cracks are perpendicular to the direction of the flow,
  i.e.\ 
  the direction of the external forcing.
Details of the experiment will be reported elsewhere~\cite{Kawazoe}.

As a first step in the theoretical investigation
  into the slope flow of $\mathrm{CaCO_3}$ paste,
  we study the dynamics
  of an elastoplastic liquid layer with constant thickness
  falling down an inclined wall.
First, we construct a continuum model equation 
  that meets several requirements,
  so that it can be a good description of the $\mathrm{CaCO_3}$ paste.
Next, we apply this model equation
  to the two-dimensional slope flow with constant layer thickness.
We find that the flow develops
  \emph{tension} in the streamwise direction,
  which remains in the paste.
Since the residual tension implies
  that the dried paste will be more fragile 
  in the pertinent direction,
  this result presents
  a possible clarification of the Type-I Nakahara effect.


\section{Requirements for the model}
\label{sec:requirements}

The strategy in this paper includes
  the construction of a set of model equations
  acceptable as a continuum description of $\mathrm{CaCO_3}$ paste.
A useful precedent for this model construction can be found
  in the continuum mechanics
  of gases and simple liquids~\cite{Landau.fluid,Schlichting.Book2000},
  in which the Navier-Stokes equation is deduced 
  from several macroscopic requirements,
  such as homogeneity, isotropy, and the postulation
  that the deviatoric stress tensor is a linear function
  of the rate-of-strain tensor (without time lag). 
Following this precedent, 
  let us list the analogous requirements for paste flows.

We assume that the dynamics of the paste is isotropic,
  in the sense that the paste has no preferred direction
  except for the principal axes of the stress tensor.
This is plausible for $\mathrm{CaCO_3}$,
  which consists
  of spherical particles~\cite{Nakahara.JPSJ74,Nakahara.JStat2006}.
On the other hand, magnesium carbonate hydroxide
  is not expected to exhibit isotropy in this sense,
  as its particles are disk-like~\cite{Nakahara.PRE74}
  and therefore can exhibit anisotropy
  similar to that of liquid crystals.

The stresses in the pastes under present consideration
  are primarily sustained by the interparticulate bond network.
There should be also a contribution
  from the viscosity of the solvent (water),
  but we assume that this contribution is much smaller 
  than that of the interparticulate bonds (in other words,
  we consider only very thick colloids).
Unlike the chemical bonds,
  the interparticulate bonds in flowing pastes
  are usually so breakable
  that they are constantly destroyed and reconstructed.
The stress is therefore expected to be governed
  by a Maxwell-type equation~\cite{Joseph.Book1990,
  Miyamoto.PRL88,Kruse.PRL92,Kruse.EurPhysJE16,Ooshida.PRL95}
  whose relaxation time represents the lifetime of the bond. 

We postulate that the relaxation time, denoted by $\tau$,
  is a scalar:
  the collapse of the force network involves
  bond breakage in all directions.
Since the paste is plastic,
  the relaxation time $\tau$ must be variable.
An infinitely large $\tau$
  represents solid-like behavior,
  while a finite $\tau$ denotes fluidity.
The transition between these two behaviors
  with a certain threshold 
  gives a formulation of plasticity.
Isotropy dictates 
  not only that $\tau$ itself is a scalar,
  but also that $\tau$ should be a function of some scalar quantity.
With the von Mises criterion~\cite{Hill.Book1950}
  and its energetic interpretation~\cite{Hencky.ZAMM4} in mind,
  we assume
  that the relaxation time $\tau$ is a function of strain energy.
Introducing $\Eps$
  to denote the nondimensionalized strain energy (defined later),
  this assumption is formulated as
\begin{equation}
  \tau = \tau(\Eps) 
       \sim \begin{cases}
	       +\infty          & (\Eps <   \text{threshold} ) \\
	       \tau_0 = \etaP/S & (\Eps \gg \text{threshold} ) 
	    \end{cases}
	    \label{*1}
\end{equation}
  where $\etaP$ is a constant with the dimension of viscosity,
  and $S$ is the shear modulus.

We describe the relaxation of the bond network
  in terms of the Lagrangian (material) variable $\BoldXi$,
  rather than the Eulerian variable $\rvect$.
The main reason for this choice
  is the adequacy of the Lagrangian description
  for tracing so-called frozen quantities.
With the relevant physical quantity
  provisionally symbolized as $\mathcal{G}$ (probably
  representative of the density of the bond network),
  the equation of relaxation is expected to have the form
\begin{equation}
 \left(1 + \tau \frac{\rD}{\rD{t}} \right) 
  \mathcal{G}(\BoldXi,t)
  = \mathcal{G}_*(\BoldXi,t)  \label{*L1}.
\end{equation}
In the limit of an infinitely long relaxation time ($\tau \to +\infty$),
  Eq.~(\ref{*L1}) reduces itself to
\begin{equation}
  \dt \mathcal{G}(\BoldXi,t) = 0   \label{*L2}
\end{equation} 
  which manifests directly
  that $\mathcal{G}$ is ``frozen'' in the material.
In the Eulerian description,
  the same assertion as Eq.~(\ref{*L2}) would have 
  a more complicated form,
\begin{equation}
  \left( \dt + \mb{v}\cdot\nabla \right) \tilde{\mathcal{G}}(\rvect,t)
   + \cdots  = 0  
   \qquad
  (\text{Eulerian})
 \label{*L3},
\end{equation} 
  where ``$\cdots$'' stands for various convective terms 
  required according to the tensorial character
  of $\tilde{\mathcal{G}}$.
Since we prefer the clarity of Eq.~(\ref{*L2})
  to the obscurity of Eq.~(\ref{*L3}),
  the Lagrangian description is adopted
  during the construction of the model (the result
  could be reformulated in the Eulerian description 
  \emph{after} it is developed, of course).

Generally, 
  the mathematical formulation of elasticity
  is related to the \emph{deformation} 
  of the fluid (or material) elements.
In the steady and quasi-steady motions of pastes,
  deformation (as opposed to \emph{rate of deformation})
  can increase unlimitedly as the time elapses.
This requires our model
  to be free from the restrictive assumption of a small deformation,
  motivating the inclusion of the geometrical nonlinearity
  to the full extent.
Besides,
  the solid behavior of the paste for a small deformation
  should have an isotropic Hookian limit (with shear modulus $S$),
  because the paste is isotropic.
For the same reason,
  the fluid behavior is expected to have a Navier-Stokes limit 
  for small $\tau$ or small shear rate (which is realized
  for water-rich pastes with a vanishingly small yield stress).
In both behaviors,
 we regard the paste as incompressible,
  as far as flow processes are concerned,
  neglecting the slow effects of drainage and evaporation.
Finally, the model equation
  must have ``relabeling symmetry''~\cite{Bennett.Book2006}, 
  i.e.\ 
  the system of equations must remain formally unchanged
  in regard to the change in the Lagrange variables.
In what follows, 
  while making some additional assumptions,
  we will construct a system of model equations 
  that satisfies all of these requirements.


\section{Model} 
\label{sec:model}

In this section,
  we construct a continuum paste model
  for a generic $\nd$-dimensional geometry.
The model equations will be summarized
  at the end of \S\ref{subsec:constitutive}.
Subsequently, in \S\ref{sec:slope-flows} and \S\ref{sec:analysis},
  this model will be analyzed under a specific setup
  describing a two-dimensional slope flow with constant layer thickness.
Readers
  who are more interested in the analysis than the model construction
  may,
  after checking Figs.~\ref{Fig:nu} and \ref{Fig:S},
  skip to Eqs.~(\ref{eqs:1D}) at the end of \S\ref{sec:slope-flows}.

\subsection{Kinematics}

First, we review 
  the Lagrangian description of kinematics.
The configuration of an $\nd$-dimensional continuum
  is represented by a mapping
  from Lagrangian variable $\BoldXi$ (also known 
  as ``label'' or ``material variable''~\cite{Bennett.Book2006})
  to the position vector $\rvect$.
For $\nd = 3$, we write
\begin{equation}
 \BoldXi = (\xi,\eta,\ze)
 \mapsto 
 \rvect = \rvect(\BoldXi,t) 
 = \begin{bmatrix} 
    x(\xi,\eta,\ze,t) \\ 
    y(\xi,\eta,\ze,t) \\ 
    z(\xi,\eta,\ze,t) 
   \end{bmatrix}_{\Cartesian}			
 \label{k1}
\end{equation}
  where
  $\left[\quad\right]_{\Cartesian}$
  denotes the representation in terms of Cartesian components.
For $\nd=2$
  we will omit $\eta$ and $y$,
  assuming that all the motion occurs in the $(x,z)$-plane.

The time-derivative of $\rvect = \rvect(\BoldXi,t)$ gives
  the velocity,
\begin{equation}
 \mb{v} = \dt\rvect(\BoldXi,t)  
 \label{k2a}.
\end{equation} 
In Eq.~(\ref{k2a}) and in what follows,
  $\dt$ stands for the time-derivative in the Lagrangian
  description (Lagrange derivative, which is usually denoted
  by $\mathrm{D}/{\mathrm{D}t}$ in Eulerian description).
Using $\left\{ \B\xi, \B\eta, \B\ze \right\}$ (where
  $\Di = \rD/{\rD{\xi^i}}$)
  as the set of local bases,
  we can represent the velocity as
\begin{equation}
 \mb{v} = v^i \B{i} \label{k2b}.
\end{equation}
In Eq.~(\ref{k2b}) and in what follows,
  summation over $i \in \{ \xi, \eta, \ze \}$ is understood
  according to Einstein's contraction rule.
The coefficients $(v^i)$ in Eq.~(\ref{k2b})
  are referred to as the contravariant components of $\mb{v}$ (see
  Eqs.~(\ref{contravariant}) and (\ref{get-comp-contravariant})
  in Appendix~\ref{app:diff}).
The acceleration is %
  $\dt\mb{v} = \dt\left( v^i \B{i} \right)$;  
  we emphasize again that $\dt$ denotes the Lagrange derivative.

The square of the Euclidean distance
  between two neighboring ``particles,''
  labeled by $\BoldXi$ and $\BoldXi + \D\BoldXi$,
  is
\begin{equation}
 \D{s}^2 = \left| {(\B{i})\,\D\xi^i} \right|^2 
  = g_{ij} \dXi^i \dXi^j , \quad
  g_{ij} = (\Di \rvect) \cdot (\Dj \rvect)
  \label{k5},
\end{equation}
  which introduces the metric tensor
  denoted by $(g_{ij})$ or $\mathsf{g}$.
In this paper
 we refer to $\mathsf{g}$ as the ``Euclidean'' metric tensor,
  which does not mean that $g_{ij}$ is equal to Kronecker's delta
  but means that
  the Euclidean metric of the $\rvect$-space
  is imported into the $\BoldXi$-space by Eq.~(\ref{k5}).

In general, it is totally unnecessary to choose $\BoldXi$
  to be some ``initial'' position of the element,
  except for some particular situations
  in which the initial state has a special significance.
One of these special cases is that of purely elastic bodies
  initially set in a stress-free and undeformed state, called
  a ``natural state''~\cite{Marsden.Book1994}.
It is meaningful in this case
  to choose the ``natural state'' position vector as $\BoldXi$  
  so that $g_{ij}$ defined by Eq.~(\ref{k5})
  is essentially identical 
  to the Cauchy-Green deformation tensor~\cite{Joseph.Book1990}
  whose difference from $\delta_{ij}$ is responsible
  for the elastic restoring force.
This is
  a rather special case, however.
More generally,
  $\BoldXi$ has nothing to do with the initial state,
  and the \emph{natural metric tensor} $\mathsf{g}^\Nat$ is used
  as a reference to define the elastic deformation,
  instead of assuming the global existence
  of the stress-free natural state.
The (locally) undeformed state
  is formulated as $\mathsf{g} = \mathsf{g}^\Nat$,
  and the difference between $\mathsf{g}$ and $\mathsf{g}^\Nat$
  is responsible for the stress.
More details about $\mathsf{g}^\Nat$
  will be discussed later.

The incompressibility condition 
  is expressed as
\begin{equation}
 \dt \detG = 0   \label{k6a},
\end{equation}
  because the mass of a fluid element is
  $\rho\sqrt{\detG}\,{\D^{\nd}\BoldXi}$
  which should remain unchanged,
  and the density $\rho$ also remains unchanged during the motion.
For simplicity, 
  we assume that $\rho$ is a global constant.
Then, without loss of generality,
  we can replace Eq.~(\ref{k6a}) by
\begin{equation}
 \detG = 1         \label{k6b}.
\end{equation}

\subsection{Equation of motion and constitutive relation}
\label{subsec:constitutive}

Now we detail
  the dynamics.
With the stress field denoted
  by $\TensorP = P^{ij}(\B{i})\otimes(\B{j})$
  and the external body force by $\mb{F} = F^i\B{i}$,
  the momentum equation is written as 
\begin{equation}
  \rho\,\dt\left( v^i \B{i} \right)
  = -\dd{}{\rvect} \cdot \Tp{\left( P^{ij} (\B{i})\otimes(\B{j}) \right)}
  + F^i \B{i}
  \notag
\end{equation}
  or, in contravariant component representation,
  as~\cite{Marsden.Book1994}
\begin{equation}
 \rho\,\left( \dt v^i + {v^j \nabla_j} v^i \right)
  = - \nabla_j P^{ij} + F^i
  \label{d1}.
\end{equation}
The left-hand side
  is the contravariant component
  of the acceleration vector $\dt\mb{v}$
  multiplied by the density $\rho$,
  and $\nabla_j$ denotes
  the covariant derivative (these mathematical concepts
  are clarified in Appendix~\ref{app:diff}
  to the degree sufficient for the present work;
  for a more profound understanding of the mathematical background,
  see Refs.~\cite{Marsden.Book1994,Nakahara.Book1990}).
While $\mb{F}$ is regarded as given,
  $\TensorP$ must be determined by a suitable constitutive relation.

From the discussion in the previous section,
  we expect that $\TensorP$ obeys
  a viscoelastic equation of Maxwell type.
The Maxwell model is often illustrated
  as a spring and dashpot connected in series~\cite{Joseph.Book1990},
  for which 
  the relation between the tension $T$ and the total length $x$
  is given by 
\begin{equation}
  T = \kappa\,\left( \xS - \xS^\Nat \right) = \mu \DD{\xD}{t}, 
  \qquad
  x = \xS + \xD  \label{d2}
\end{equation}
  where $\kappa$ is the spring constant,
  $\mu$ is the resistance, 
  $\xS$ and $\xD$ are
  the length of the spring part and the dashpot part,
  respectively,
  and $\xS^\Nat$ denotes the natural length of the spring part.
It is customary to 
  eliminate the ``internal'' variables ($\xS$, $\xD$ and
  $\xS^\Nat$) from Eq.~(\ref{d2}),
  which yields
\begin{equation}
 \left(\mu^{-1} + \kappa^{-1} \DD{}{t} \right) T = \DD{x}{t}
  \label{d3}.
\end{equation}
A timescale $\mu/\kappa$ in regard to stress relaxation
  is recognized in Eq.~(\ref{d3}).

Now it is necessary to elaborate the Maxwell model
  in two respects:
  it needs to include plasticity
  and it also needs to describe $\nd$-dimensional continuum mechanics.
In regard to the first point,
  most of the existing studies
  are based on an elasto-plastic decomposition,
  which is a direct extension of Eq.~(\ref{d2}).
However, this approach has a disadvantage in
  that the incautious use of internal variables can lead to a difficulty,
  in particular
  for a finite deformation~\cite{Lee.ASME-JAM36,Lee.ASME-JAM48}.
Here we adopt a different approach 
  that is closer to Eq.~(\ref{d3}),
  thereby avoiding a direct reference to the internal variable $\xD$.

The essential idea is to attribute the relaxation
  to the natural length $\xNat$,
  which is related to the tension $T$
  as if the model is totally elastic:
\begin{equation}
 T = \kappa\,(x - \xNat)   \label{d4}.
\end{equation}
The natural length $\xNat$ can be expressed 
  as $\xNat = \xS^\Nat+\xD^{}$ in terms of internal variables,
  but this relation is not to be used explicitly;
  we note only that $\xNat$ is time-dependent 
  while $\xS^\Nat$ is not.
By substituting Eq.~(\ref{d4}) into Eq.~(\ref{d3}),
  we find an equation that describes
  the relaxation of the natural length $\xNat$:
\begin{subequations}%
 \begin{equation}
  \DD{\xNat}{t} = {\frac{\kappa}{\mu}}\left( x - \xNat \right)
   \label{d5a},
 \end{equation}
 or, by introducing $\tau = \mu/\kappa$, as
 \begin{equation}
  \left( 1 + \tau\,\DD{}{t} \right) \xNat = x   \label{d5b}
 \end{equation}%
 in the form of relaxation toward $\xNat = x$.
 \label{eqs:d5}%
\end{subequations}%
Eqs.~(\ref{d4}) and (\ref{eqs:d5})
  provide us with a prototype of the plastic model.

Let us find $\nd$-dimensional continuum equations
  corresponding to the prototypical
  equations (\ref{d4}) and (\ref{eqs:d5}).
As a candidate, we adopt an elastic constitutive equation
\begin{equation}
 P^{ij} 
  = \tilde{p} g^{ij}
  + S\,\left( g^{ij} - g_\Nat^{ij} \right)  \label{d6}
\end{equation}
  together with an inelastic equation
\begin{equation}
 \dt{g_\Nat^{ij}} = -\nu {g_\Nat^{ij}} + \nu_* {g^{ij}}  \label{r1},
\end{equation}
  where $(g^{ij})$ 
  denotes the inverse of the component matrix
  of the ``Euclidean'' metric tensor $(g_{ij})$,
  and $(g_\Nat^{ij})$ is that of the natural metric tensor,
  such that
\begin{equation}
 {g_{ij}}      {g^{jk}}      = 
 {g^\Nat_{ij}} {g_\Nat^{jk}} = {\delta_i}^k 
  \notag.
\end{equation}

The natural metric tensor $\mathsf{g}^\Nat$ 
  represents the square of the ``natural distance''
  between two neighboring points 
  labeled by $\BoldXi$ and $\BoldXi + \D\BoldXi$,
\begin{equation}
  \left({\D{s}^\Nat}\right)^2
  = g^\Nat_{ij} \dXi^i \dXi^j  \label{e1},
\end{equation}
  in the sense that the difference
  between $\D{s}^2$ and $\left({\D{s}^\Nat}\right)^2$
  accounts for the restoring force according to Eq.~(\ref{d6}).
In the special case of purely elastic bodies
  initially set in a stress-free ``natural state'' (at $t=t_0$),
  $\D{s}^\Nat$ is the distance in this initial configuration
  and $\mathsf{g}^\Nat$ is the corresponding metric:
\begin{equation}
  g^\Nat_{ij} = \Ev{g_{ij}}{t=t_0}  \qquad
  (\text{purely elastic case})  \notag.
\end{equation}
In general, however,
  $\mathsf{g}^\Nat$ differs from the initial value of $\mathsf{g}$.
This is inevitable due to Eq.~(\ref{r1}), 
  which prescribes
  that the natural metric $\mathsf{g}^\Nat$ is subject to relaxation.  
To make Eq.~(\ref{r1})
  more easily recognizable as a relaxation equation,
  we rewrite it in the manner of Eqs.~(\ref{*L1}) and (\ref{d5b}) as 
\begin{equation}
 \left( 1 + \tau\,\dt \right) g_\Nat^{ij}
  = K g^{ij},
 \quad
 \tau = \nu^{-1},     \quad
 K = \frac{\nu_*}{\nu}          \label{d7};
\end{equation}
  this equation provides that $(g_\Nat^{ij})$ should evolve
  toward an isotropic tensor $(K g^{ij})$.
Plasticity is incorporated via $\tau$
  according to Eq.~(\ref{*1}).
The idea of using a natural metric
  to reformulate the Maxwell model has been known 
  among several researchers of rheology (including 
  the authors of Refs.~\cite{Kruse.PRL92,Kruse.EurPhysJE16}),
  but the present author could not identify any publications
  in which the notion of the natural metric and its relaxation
  is formulated explicitly.

The $\nd$-dimensional elastic equation~(\ref{d6}),
  corresponding to the
  one-dimensional Hookian equation~(\ref{d4}),
  originates from consideration of elastic strain energy.
Since $ g^\Nat_{ij} \dXi^i \dXi^j $
  is a positive definite quadratic form,
  there exists a set of Euclidean vectors
  $\{ \mb{p}_\xi, \mb{p}_\eta, \mb{p}_\ze \}$
  such that $g^\Nat_{ij} = {\mb{p}_i} \cdot {\mb{p}_j}$ (this is
  proved essentially in the same way as the polar decomposition
  theorem~\cite{Joseph.Book1990,Marsden.Book1994}).
Then,
  by defining
\begin{equation}
 \DrNat = \mb{p}_i \,\dXi^i   \label{e2},
\end{equation}
  we have
  $\left({\D{s}^\Nat}\right)^2 
  = \DrNat\cdot\DrNat = \left| \DrNat \right|^2$.
Note that Eq.~(\ref{e2}) does not claim
  that $\DrNat$ is a differential of ``$\rvect^\Nat$'':
  such integrability is not guaranteed.
However,
  it is legitimate to interpret $\DrNat$
  as a natural configuration of each small element.
Since $\mb{p}_i$'s must be linearly independent
  due to the positivity of $\detGNat$,
  Eq.~(\ref{e2}) can be inverted,
  which we denote as
  $\dXi^i = {\mb{p}_*^i} \cdot \DrNat$.
From this and the ``Euclidean'' metric~(\ref{k5}),
  we have a relation between the Euclidean distance $\D{s}$
  and the natural configuration $\DrNat$,
\begin{equation}
 \D{s}^2 
  = \left( 
     g_{ij}\, {\mb{p}_*^i} \otimes {\mb{p}_*^j} 
   \right) : \left( {\DrNat \otimes \DrNat} \right)
  \label{e3}.
\end{equation}
Let us denote the eigenvalues of this quadratic form 
  by $\{ \ld_\alpha^2 \}$
  so that 
  $ \D{s}^2 = \ld_\alpha^2 \left|{\DrNat}\right|^2 $
  along the $\alpha$-th principal axis.
The geometrical meaning of $\ld_\alpha$ is clear:
  it represents the elongation factor of the line element.
Isotropy requires
  that the elastic energy (denoted by $E$) should consist
  of a symmetric combination of these eigenvalues.
The simplest form with a correct Hookian limit 
  is
\begin{equation}
 E = {\frac12}S \left( \ld_1^2 + \ld_2^2 + \ld_3^2 - 3 \right)
 \label{e4}
\end{equation}
  for $\nd=3$.
Eq.~(\ref{e4}) is known
  as neo-Hookian constitutive equation~\cite{Marsden.Book1994}.
With the aid of the incompressibility condition,
  which implies $ \ld_1 \ld_2 \ld_3 = 1 $,
  Eq.~(\ref{e4}) reduces to 
  $E = S (e_1^2 + e_2^2 + e_3^2)$
  for small deformations ($\ld_\alpha = 1 + e_\alpha$ and 
  $|e_\alpha| \ll 1$).
By using the definition of $\mb{p}_*^i$
  and introducing 
  $\Eps = \sum_\alpha \left( \ld_\alpha^2 - 1 \right)$, 
  for finite deformations,
  the elastic energy $E$ is expressed
  in terms of the inverse natural metric tensor:
\begin{equation}
 E = {\frac12}S\Eps, \quad  
  \Eps = g_{ij} g_\Nat^{ij} - \nd   \label{e5}.
\end{equation}

By calculating the variation of the elastic energy $E$
  in regard to $\rvect$ through the metric tensor $\mathsf{g}$
  under the constraint of incompressibility condition~(\ref{k6b}),
  we find that the contravariant components of the stress tensor
  are given by Eq.~(\ref{d6}).
Details of this calculation
  are shown in Appendix~\ref{app:var-E}.
Note that the tensor $(g^{ij})$ 
  in the first term of the right-hand side of Eq.~(\ref{d6})
  stands for the Euclidean unit tensor:
\begin{equation}
 g^{ij}(\B{i})\otimes(\B{j}) = \openone  \label{k7}.
\end{equation}
Thus we find 
  that the term $\tilde{p}\,g^{ij}$ stands for an isotropic stress.
The scalar $\tilde{p}$ is related to the hydrostatic pressure
  arising as a constraint force (Lagrange multiplier)
  for incompressibility.
It is convenient to define 
\begin{equation}
 \TensorSigma 
 = S\,\left( g_\Nat^{ij} (\B{i})\otimes(\B{j}) - \openone \right) 
 \label{e6}
\end{equation}
  and call it the ``elastic stress tensor''
  so that the stress tensor
  $\TensorP = P^{ij}(\B{i})\otimes(\B{j})$
  is given by
\begin{equation}
 \TensorP     
  = \tilde{p} \openone - \TensorSigma   \label{e6+}.
\end{equation}
It is easy to confirm that
  $\TensorSigma$ vanishes when $g^\Nat_{ij} = g_{ij}$.

We emphasize that $(g_\Nat^{ij})$ in Eq.~(\ref{e6}),
  which determines the elastic stress tensor $\TensorSigma$,
  is the \emph{inverse} of the natural metric tensor.
This must be the case
  so that $\TensorSigma$ should remain invariant
  under the relabeling of the Lagrange variables.
This is also acceptable
  if we remember that springs with different lengths
  but the same local properties
  obey a constitutive relation analogous to Eq.~(\ref{e6}),
\begin{equation}
 T = s_0\,\left( \frac{x}{x^\Nat} - 1 \right)   \notag
\end{equation}
  where $s_0$ is the normalized spring constant,
  and notice that $\TensorSigma$ is
  an intensive variable as well as the tension $T$
  and therefore must be expressed as such.

Now we discuss the inelastic part of our model
  described by Eq.~(\ref{r1}).
This equation states 
  the relaxation of the inverse natural metric tensor,
  formulated according to the following discussion
  on interparticulate bonds.
The natural metric represents
  the energetically optimal configuration of the particles
  determined by the bond network.
The network strength, or the bond density,
  is represented by the inverse natural metric $\mathsf{g}_\Nat$
  (not by the natural metric $\mathsf{g}^\Nat$ itself).
In flowing pastes, however, 
  this bond network is ephemeral. 
We suppose that the network is destroyed at some rate
  and reconstructed isotropically.
With the destruction rate denoted by $\nu$
  and the reconstruction rate by $\nu_*$,
  the temporal change of the bond density
  is given by $-\nu {g_\Nat^{ij}} + \nu_* {g^{ij}}$, 
  leading to Eq.~(\ref{r1}).

The ratio $K = \nu_*/\nu$ is determined
  by postulating the incompressibility of $\mathsf{g}^\Nat$,
\begin{equation}
 \detGNat = 1  \label{r2}.
\end{equation}
Differentiating Eq.~(\ref{r2}) with regard to $t$
  and then substituting Eq.~(\ref{r1}) into it,
  we find that $\nd\nu = {g^\Nat_{ij}}{g^{ij}}\nu_*$,
  which implies
\begin{equation}
 K = \frac{\nd}{g^\Nat_{ij} g^{ij}} 
  \qquad
  \left( \,\text{i.e.}\quad
   \nu_* = \frac{\nd}{g^\Nat_{ij} g^{ij}} \,\nu 
 \right)
 \label{r3}.
\end{equation}

\begin{figure}
 \includegraphics[clip,width=5.5cm]{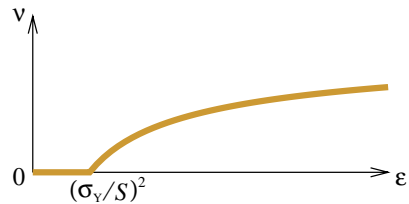}  
 \caption{\label{Fig:nu}%
 The inverse of the relaxation time,
 $\tau^{-1} = \nu(\Eps)$, defined by Eq.~(\ref{*3}).
 According to neo-Hookian constitutive equation
 of $\nd$-dimensional elastic bodies,
 $\Eps$ is given by Eq.~(\ref{e5}).
 }%
\end{figure}

According to Eq.~(\ref{*1}) in the previous section,
  $\tau$ is supposed to be a function of the elastic strain energy,
  so that $\tau = \tau(\Eps)$
  with $\Eps$ given by Eq.~(\ref{e5}).
The simplest form
  consistent with Eq.~(\ref{*1}) is
\[
  \tau = \begin{cases}
           +\infty  &  (\Eps < \Yield^2/S^2) \\
            \etaP/S &  (\Eps > \Yield^2/S^2) 
         \end{cases}
\]
  where $\Yield$ is the yield stress (we will see later
  that the energy $\Eps$ for shear stress $\sigma$
  is calculated to be $\sigma^2/S^2$).
It is physically more realistic
  and mathematically less problematic
  to suppose that $\tau$ is a continuous function of $\Eps$.
Here we assume
\begin{equation}
 \tau = \nu^{-1},
 \quad
 \nu = \nu(\Eps) 
 = {\frac{S}{\etaP}} \max\left(
			  0,\; 
			  1-\frac{\sigma_{\text{Y}}/S}{\sqrt{\Eps}}
			\right)
 \label{*3}
\end{equation}
  which is Lipschitz-continuous 
  in spite of weak singularity at the yield point (Fig.~\ref{Fig:nu}).
Eq.~(\ref{*3}) is chosen in such a way that it agrees with 
  Bingham plasticity~\cite{Bingham.Book1922,Mei.JFM431,Ooshida.PRL95}
  for simple shear flow with shear rate $\dot\gamma$,
  where the shear stress $\sigma$ is estimated to be 
  $\sigma \simeq S\tau{\dot\gamma}$.
Admitting $\Eps \simeq \sigma^2 / S^2$,
  from Eq.~(\ref{*3}) we find
\[
 \dot\gamma \simeq \frac{\nu\sigma}{S} = \nu(\Eps)\,\sqrt{\Eps}
    = \begin{cases}
       0                                    & (\Eps < \Yield^2/S^2) \\
       \etaP^{-1} ( S\sqrt{\Eps} - \Yield ) & (\Eps > \Yield^2/S^2) 
      \end{cases}
\]
  which is Bingham plasticity.

Let us summarize our model.
The governing system of equations
  consists of Eqs.~(\ref{d1}), (\ref{d6}), (\ref{d7}),
  and (\ref{*3}),
  supplemented with the kinematic relations 
  (\ref{k2a}), (\ref{k2b}) and (\ref{k5}), 
  as well as incompressibility
  conditions (\ref{k6b}) and (\ref{r2}).
Eq.~(\ref{*3}) requires the evaluation of $\Eps$ by Eq.~(\ref{e5}),
  which is actually not independent of Eq.~(\ref{d6}),
  but should be included in the model for convenience.
The independent variables are $\BoldXi$ and $t$ (Lagrangian 
  description), 
  and the essential dependent variables
  are $\rvect$ and $\mathsf{g}^\Nat$.
The velocity and the Euclidean metric tensor
  are derived from the differentials of $\rvect = \rvect(\BoldXi,t)$.
Due to the incompressibility condition,
  there arise two additional scalar fields,
  namely $\tilde{p}$ and $K$;
  the latter is determined by Eq.~(\ref{r3}).


\subsection{Navier-Stokes limit}
\label{subsec:NS}

There remains the task
  to confirm that the whole system of model equations
  reduces to the $\nd$-dimensional incompressible Navier-Stokes equation
  if $\tau$ is set to be a small constant
  such that $\tau \ll \|{\nabla\mb{v}}\|^{-1}$.
By expanding $g^\Nat_{ij}$ (as well as $g_\Nat^{ij}$) and $K$
  in power series of $\tau$,
  from Eqs.~(\ref{d7}) and (\ref{r2})
  we find 
\begin{equation}
 g_\Nat^{ij} = g^{ij} - \tau\dt{g^{ij}} + O(\tau^2),  \quad
  K = 1 + O(\tau^2)  
  \label{*x}.
\end{equation}
The time-derivative term $\dt{g^{ij}}$
  on the right-hand side of Eq.~(\ref{*x})
  is calculated as
\begin{subequations}%
 \begin{equation}
  \dt{g^{ij}} = -g^{ik} g^{jl} \dt{g_{kl}}
 \end{equation}
 and
 \begin{align}
  \dt{g_{ij}} 
  &= \dt\left( (\B{i})\cdot(\B{j}) \right)  \notag\\
  &=
  (\Di\mb{v})\cdot(\B{j}) + (\B{i})\cdot(\Dj\mb{v})   \notag\\
  &= \nabla_i {v_j} + \nabla_j {v_i}
 \end{align}%
 \label{eqs:dt.g}%
\end{subequations}%
  where $(v_i)$ denotes the covariant components
  of the velocity vector $\mb{v}$,
  and $\nabla_i{v_j}$ denotes the covariant derivative of $v_j$
  defined by
  $ \Di (v_j \nabla{\xi^j}) = (\nabla_i{v_j}) \nabla{\xi^j} $.
Using Eqs.~(\ref{eqs:dt.g})
  to evaluate $g_\Nat^{ij}$ in Eq.~(\ref{*x}),
  from Eq.~(\ref{e6}) we obtain
\begin{align*}
 \TensorSigma 
 &= S\,\left( g_\Nat^{ij} (\B{i})\otimes(\B{j}) - \openone \right) 
 \notag \\
 &= S\,\left( g_\Nat^{ij} - g^{ij} \right) (\B{i})\otimes(\B{j})
 \notag\\
 &= - S\tau \left( \dt{g^{ij}} \right) (\B{i})\otimes(\B{j})
 \notag\\
 &= S\tau \left( g^{ik} g^{jl} \dt{g_{kl}} \right)
 (\B{i})\otimes(\B{j})                                \notag\\
 &= S\tau\, (\dt{g_{kl}}) (\nabla{\xi^k})\otimes(\nabla{\xi^l})
 \notag\\
 &= S\tau\, \left(  
 \nabla_k {v_l} + \nabla_l {v_k}
 \right) (\nabla{\xi^k})\otimes(\nabla{\xi^l}) 
 \notag\\
 &= S\tau\,\left( 
 \nabla\otimes\mb{v} + \Tp{(\nabla\otimes\mb{v})} 
 \right),
\end{align*}
  and identify it with the Newtonian-Stokesian relation
\begin{equation}
 \TensorSigma 
 = 2 \eta_* \,\mathrm{sym}\,\mathrm{grad}\,\mb{v} \qquad
 \left( \eta_* = S\tau \right)  
 \label{NS}
\end{equation}
  where $\sym\grad\mb{v}$ denotes
  the symmetric part of $\nabla\otimes\mb{v}$.  
The equation of motion
  then reduces to the Navier-Stokes equation,
  which was to be demonstrated.


\begin{figure}
 \includegraphics[clip,width=5.0cm]{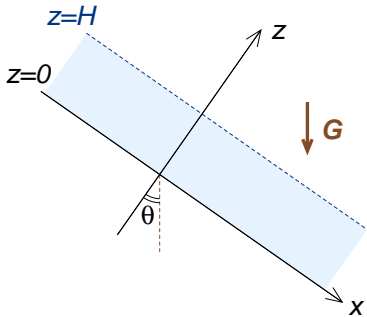}  
 \caption{\label{Fig:S}%
 Schematic view of the system and coordinates.
 A uniform fluid layer with thickness $H$ is assumed.
 The label variable $\ze$
 coincides with the depthwise Eulerian coordinate $z$ 
 in the present setup,
 and the velocity is $U{\mb{e}_x}$ 
 where $U = \dt{X(\ze,t)}$.
 The gravitational acceleration vector 
 is $\mb{G} = (G\sin\theta){\mb{e}_x} - (G\cos\theta){\mb{e}_z}$,
 whose $x$-component is $G_x = G\sin\theta$.
 }%
\end{figure}


\section{Simplification for slope flows with uniform thickness}
\label{sec:slope-flows}

We have obtained a system of equations that is
  acceptable as a model of isotropic pastes.
Next, let us analyze this system under a particular setup
  describing slope flows.
Though the equations are $\nd$-dimensional
  and it is also possible to formulate boundary conditions
  for fully three-dimensional surface deformations,
  it is not wise trying to solve the full system immediately
  by direct numerical simulations,
  as it would require too much difficulty and provide too little insight.
Rather, the first thing to do
  is to elucidate the basic behavior of the model
  in the simplest situation.

Here, we study a two-dimensional paste flow ($\nd=2$) 
  on a slope inclined by an angle $\theta$.
The setup of the system
  is shown in Fig.~\ref{Fig:S}.
All of the motion is supposed to occur in the $(x,z)$-plane,
  and it is in this plane
  that the paste is assumed to be isotropic.
The flow is driven 
  by the gravitational force $\mb{F} = \rho\mb{G}$,
  where $\mb{G}$ is the gravitational acceleration vector,
\begin{equation}
  \mb{G} = 
  \begin{bmatrix}
   \phantom{+} G\sin\theta \\ -G\cos\theta
  \end{bmatrix}_{\Cartesian} 
  = (G\sin\theta)\mb{e}_x - (G\cos\theta)\mb{e}_z
  \label{Gravity}.
\end{equation}

The free surface requires 
  the dynamical boundary condition and the kinematic boundary condition.
The dynamical boundary condition prescribes
  the continuity of the stress,
  while the kinematic boundary condition postulates
  that the surface must move together with the adjacent fluid
  to satisfy the mass conservation law.
Since we assume here
  that the paste layer has a constant thickness $H$,
  the dynamical boundary condition
  reduces to $P^{ij}\nabla_j{z} = 0$ (see
  Eq.~(\ref{BC.TensorP}) and the text below it
  in Appendix~\ref{app:diff}).
The kinematic boundary condition is trivially satisfied
  by assuming 
  that the velocity field is also uniform in regard to $x$
  and parallel to the $x$-axis.

Under the present assumptions,
  the fluid motion is expressed in terms of a single function
  which we denote by $X = X(\ze,t)$, as
\begin{equation}
 \rvect 
  = 
  \begin{bmatrix}
   \xi+X \\ \ze 
  \end{bmatrix}_{\Cartesian}
  = (\xi+X)\,\mb{e}_x + \ze \mb{e}_z
  \label{u1}.
\end{equation}
The time-derivative of Eq.~(\ref{u1}) 
  gives the velocity,
\begin{equation}
\mb{v} = \dt\rvect = U\mb{e}_x
 \label{u2},
\end{equation}
  where $U = \dt{X}$.
By substituting Eq.~(\ref{u1}) into Eq.~(\ref{k5}),
  we obtain the Euclidean metric tensor $g_{ij}$ 
  expressed in terms of $X$,
\begin{equation}
 \mathsf{g} 
  = 
  \begin{bmatrix}
   g_{\xi\xi} & g_{\xi\ze} \\
   g_{\ze\xi} & g_{\ze\ze}   
  \end{bmatrix}
  =
  \begin{bmatrix}
   1  &  X'  \\
   X' &  1 + {X'}^2
  \end{bmatrix}
  \label{u3},
\end{equation}
  where $X'$ is an abbreviation for $\rD{X}/\rD\ze$.
The incompressibility condition (\ref{k6b}) is
  already satisfied
  and there is no need to require it particularly.

Let us concretize the equations containing the natural metric.
The calculation can be performed in
  at least two different ways: 
  one may evaluate the terms in the momentum equation~(\ref{d1})
  either on the ground of modern differential geometry
  of the Riemannian manifold determined by $g_{ij}$,
  or fully utilizing the Cartesian components
  in the embedding Euclidean space,
  as is shown in the latter half of Appendix~\ref{app:diff}.
Both methods yield the same result.

As the natural metric tensor for the present case
  is a $2\times2$ symmetric tensor
  with $\detGNat$ fixed to be unity,
  it can be expressed by two parameters.
We set
\begin{equation}
  \mathsf{g}^\Nat 
  = 
  \begin{bmatrix}
   g^\Nat_{\xi\xi} & g^\Nat_{\xi\ze} \\
   g^\Nat_{\ze\xi} & g^\Nat_{\ze\ze} 
  \end{bmatrix}
  = 
  \begin{bmatrix}
   e^{-\alpha} &  \beta   \\
   \beta      &  (1+\beta^2) e^\alpha
  \end{bmatrix}
  \label{g1}
\end{equation}
  with $\alpha = \alpha(\ze,t)$, $\beta = \beta(\ze,t)$,
  and calculate its inverse matrix $\mathsf{g}_\Nat$.
Then we substitute it,
  together with $\mathsf{g}^{-1}$ calculated from Eq.~(\ref{u3}),
  into the equations composing the constitutive relation.
Eq.~(\ref{d6}) then yields
  the stress tensor $\TensorP$.
Its Cartesian representation, 
  calculated from Eqs.~(\ref{e6}) and (\ref{e6+}), reads
\begin{align}
 \TensorP
  &= \tilde{p}\openone - \TensorSigma  \notag \\
  &= \tilde{p}\openone 
  - S \begin{bmatrix}
       {e^\alpha}\left( 1+\tilde\sigma^2 \right) - 1  & 
       \tilde\sigma  \\ 
       \tilde\sigma  &  e^{-\alpha} - 1
      \end{bmatrix}_{\Cartesian}
      \label{*P}
\end{align}
  where $\tilde\sigma = \sigma_{xz}/S$
  stands for the nondimensionalized shear stress,
  and is given by
\begin{equation}
 \tilde\sigma = e^{-\alpha} X' - \beta  \label{sigma}.
\end{equation}

The momentum equation~(\ref{d1})
  reads
\begin{equation}
 \rho\dt{U} 
  = S\Dz{\tilde\sigma} + \rho G_x
 \label{m1}
\end{equation}
  where $G_x = G\sin\theta$ is the $x$-directional component
  of the gravitational acceleration vector $\mb{G}$,
  given by Eq.~(\ref{Gravity}).
Note that the depthwise component of the equation of motion 
  does not participate in the dynamics,
  as it determines only the hydrostatic pressure.
  
From Eq.~(\ref{r1}) or (\ref{d7}),
  taking Eq.~(\ref{r3}) into account
  and using $\mathsf{g}$ parametrized as Eq.~(\ref{u3})
  and $\mathsf{g}^\Nat$ as Eq.~(\ref{g1}),
  we obtain
\begin{align}
 \tau \dt{\alpha} &=  1 - \frac{2}{2+\Eps} e^\alpha  \label{g2} \\
 \tau \dt{\beta}  &=  -\beta + \frac{2}{2+\Eps} X'   \label{g3}  
\end{align}
  where we have utilized the relation $K = 2/(2+\Eps)$ 
  with $\Eps$ defined by Eq.~(\ref{e5}),
  which holds for the two-dimensional case (we note
  that the three-dimensional case is not so simple).
By calculating $\Eps$ from Eq.~(\ref{e5})
  and then rewriting the result in terms of $\tilde\sigma$,
  we find
\begin{align}
 \Eps
 &= {e^\alpha}\left( e^{-\alpha}X' - \beta \right)^2 
 + 2\,(\cosh{\alpha}-1)  \notag \\
 &= {e^\alpha}{\tilde\sigma}^2 + 2\,(\cosh{\alpha}-1)  \label{Eps}.
\end{align}
Note that Eq.~(\ref{Eps}) endorses
  the relation between $\Eps$ and ${\tilde\sigma}$
  stated several lines before Eq.~(\ref{*3}),
  as long as $\alpha = o({\tilde\sigma})$
  (which is usually the case).

Though the above equations constitute a closed system,
  $X'$ and $\beta$ are inconvenient variables
  as they increase unboundedly as time elapses.
To avoid this inconvenience,
  we rewrite the equations in terms of $U$ and $\tilde\sigma$.
Using the evolution of $\tilde\sigma$
  instead of Eq.~(\ref{g3}) for $\beta$, 
  and also rewriting $\tau$ in terms of $\nu(\Eps)$,
  we obtain a system of three equations
  governing three variables,
  namely $\alpha(\ze,t)$, $\tilde\sigma(\ze,t)$, and $U(\ze,t)$:
\begin{subequations}
 \begin{align}
  \dt{\alpha} 
  &= \nu(\Eps) \left( 1 - \frac{2\,{e^\alpha}}{2+\Eps} \right) 
  \label{*a}, \\
  \dt{\tilde\sigma} 
  &= e^{-\alpha} \Dz{U} - \nu(\Eps)\,\tilde\sigma  
  \label{*s}, \\
  \dt{U}
  &= {\frac{S}{\rho}} \Dz{\tilde\sigma} + G_x  
  \label{*u}.
 \end{align}%
 \label{eqs:1D}%
\end{subequations}%
Aside from the curious equation~(\ref{*a}) for $\alpha$,
  this system of equations has a familiar form
  that can be recognized
  as a description of a slope flow (Fig.~\ref{Fig:S}),
  with Eq.~(\ref{*s}) relating 
  the nondimensional shear stress $\tilde\sigma$
  to the shear rate $\Dz{U}$,
  and Eq.~(\ref{*u}) describing momentum balance.
Plasticity is introduced via $\nu(\Eps)$
  that is the inverse
  of the relaxation time mentioned in Eq.~(\ref{*1}).
The functional form of $\nu(\Eps)$
  is specified by Eq.~(\ref{*3}) and Fig.~\ref{Fig:nu}
  on the basis of Bingham plasticity.
The nondimensional strain energy $\Eps$, defined by Eq.~(\ref{e5}),
  is evaluated 
  as a function of $\tilde\sigma$ and $\alpha$ as in Eq.~(\ref{Eps}).

Eqs.~(\ref{eqs:1D}) require
  two boundary conditions.
We pose a no-slip boundary condition at the wall,
\begin{equation}
  \Ev{U}{\ze=0} = 0  \label{BC1},
\end{equation}
  while the free-surface condition, for the present case, gives
\begin{equation}
 \Ev{\tilde\sigma}{\ze=H} = 0  \label{BC2}.
\end{equation}


\section{Analysis}
\label{sec:analysis}

\subsection{Qualitative consideration}
\label{subsec:qualitative}

Eqs.~(\ref{eqs:1D}) together with two boundary conditions
  define a closed system of evolutional equations.
The energy is supplied by gravitational work ${\rho}{G_x}U$,
  stored as elastic energy $\Eps$,
  and dissipated through the relaxation of $\mathsf{g}^\Nat$
  that represents the viscous part of the Maxwell model.
Plasticity implies
  that the dissipation process is limited by a threshold,
  in such way
  that the relaxation time can become infinitely large
  according to Eq.~(\ref{*3}).
This allows some part of the elastic energy
  to remain frozen inside the paste.

In the present study,
  $\alpha$ plays an important role. 
Eq.~(\ref{*a}) clarifies
  that the threshold mechanism included in $\nu(\Eps)$,
  shown in Fig.~\ref{Fig:nu},
  governs the fundamental behavior of $\alpha$.
For an $\Eps$ smaller than the threshold value,
  $\nu(\Eps)$ vanishes
  and therefore a practically arbitrary function of $\ze$ is admissible
  as a steady solution to Eq.~(\ref{*a}),
  as long as it allows $\Eps$ to stay within the threshold.
This implies a strong non-uniqueness of $\alpha$
  that can remain in the static paste;
  there are an infinitely large number of possibilities,
  whose realization depends on the time-dependent process
  of evolution (an analogous situation occurs
  also in dry granular materials
  subject to static friction~\cite{Duran.Book2000}).
On the other hand,
  $\alpha$ in the flowing paste is expected to evolve
  toward a steady solution that is uniquely determined
  if the external force, film thickness and paste properties
  are specified.
This steady solution will be provided later in a closed form.

We will show
  that the residence of an $\alpha > 0$ in the paste
  means the presence of $x$-directional tension.
Then we will derive a steady solution for a flowing paste analytically,
  showing that $\alpha$ is positive there.
Time-dependent numerical calculations for flowing pastes
  typically exhibit relaxation toward this solution,
  involving the creation of a positive $\alpha$.
The numerical calculations also show
  that some portion of $\alpha$ remains in the paste
  after its flow is stopped,
  and the residual value of $\alpha$ is still positive.
This process creates an $x$-directional tension
  remaining in the paste
  and therefore gives a possible clarification 
  of the Type-I Nakahara effect.

\subsection{Residual tension}

Let us confirm
  that $\alpha>0$ implies tension.
This is intuitively evident
  if we recall that $e^{-\alpha}$
  stands for the $\xi\xi$-component
  of the natural metric tensor~(\ref{g1}),
  and conceive of $e^{-\alpha} < 1$ as contraction of natural length
  of the (supposed) ``springs'' in the $x$-direction.
More formally,
  this is demonstrated by calculating
  the normal stress difference for the ``ground state''
  that minimizes the elastic energy $\Eps$ 
  as a function of $\mathsf{g}$,
  with $\mathsf{g}^\Nat$ being fixed.
In terms of the parametrization
  given by Eqs.~(\ref{u3}) and (\ref{g1}),
  the problem is to minimize $\Eps = \Eps(X',\alpha,\beta)$
  for fixed values of $(\alpha,\beta)$.

From Eq.~(\ref{Eps})
  we find that the minimizer of $\Eps(X',\alpha,\beta)$
  is $X' = {e^\alpha} \beta$,
  or equivalently $\tilde\sigma = 0 $ (vanishing shear stress).
Then, 
  using Eq.~(\ref{*P}) to calculate
  the diagonal components of $\TensorSigma$ in Eq.~(\ref{e6}),
  we find the normal stress difference
\begin{equation}
 \sigma_{xx} - \sigma_{zz} = 2 S \sinh{\alpha}    \label{*n}
\end{equation}
  for $\tilde\sigma = 0$.
Clearly, this is positive for $\alpha>0$,
  showing a residual tension in the $x$-direction.

\subsection{Steady solution for flowing pastes}

Eq.~(\ref{*n}) shows
  that a paste layer left in the unloaded state ($\tilde\sigma = 0$)
  bears an $x$-directional tension if $\alpha>0$.
The next task is to show
  that the flow makes $\alpha>0$
  if it approaches a steady solution of Eqs.~(\ref{eqs:1D}).

For steady flows,
  the nondimensional shear stress $\tilde\sigma = \sigma_{xz}/S$
  is determined by the momentum balance (\ref{*u})
  and the free surface boundary condition (\ref{BC2}).
The result is
\begin{equation}
 \tilde\sigma = {\frac{\rho G_x}{S}} (H-\ze)  \label{z1}.
\end{equation}
Note that Eq.~(\ref{z1}) holds
  for \emph{static} states as well.
For that case,
  the steady solution consists of Eq.~(\ref{z1}), $U=0$,
  and an arbitrary $\alpha=\alpha(\ze)$ 
  such that $\Eps < \Yield^2 / S^2 $ (i.e.~$\nu(\Eps)=0$).
On the other hand,
  $\nu(\Eps)$ must be non-zero for \emph{flowing} pastes,
  which makes the steady solution totally different.
For steady flows ($\nu(\Eps) \ne 0$ and $\dt{\alpha} = 0$),
  Eq.~(\ref{*a}) yields
\begin{equation}
 \Eps = 2 (e^\alpha - 1)   \label{z2}.
\end{equation}
Since $\Eps$ must be positive according to Eq.~(\ref{Eps}),
  from the above equation~(\ref{z2}) it follows 
  that $\alpha$ must be positive as well.
More concretely,
  from Eqs.~(\ref{Eps}), (\ref{z1}) and (\ref{z2})
  we find
\begin{align}
 \alpha 
 &= -{\frac12} 
 \log\left( 1 - \tilde\sigma^2 \right) \notag\\
 &= -{\frac12} 
 \log\left[\,
 1 - \left( \frac{\rho G_x}{S} (H-\ze) \right)^2
 \;\right]
 \label{z3}     
\end{align}
  for the flowing part of the paste in steady state.
It is also confirmed 
  that $\Eps \simeq {\tilde\sigma}^2$ for small $|\tilde\sigma|$.

The neighborhood of the free surface requires
  a separate treatment,
  because this region remains solidified
  due to the lack of a sufficient shear stress.
The boundary between the solidified and fluidized regions
  can be calculated by using Eqs.~(\ref{z2}) and (\ref{z3}),
  which give $\Eps = \Eps(\alpha(\ze))$ in the fluidized region,
  to find the location $\zYield$
  such that $\Eps(\alpha(\zYield)) = \Yield^2 / S^2$.
In the region $\zYield < \ze < H$
  where the paste is solidified,
  the velocity $U$ is uniform.
The velocity $U$ in the fluidized region 
  can be obtained by integrating Eq.~(\ref{*s})
  under the boundary condition~(\ref{BC1}).

As is evident from Eq.~(\ref{z1}),
  the maximum of the shear stress $\sigma_{xz}$ occurs at the wall.
The wall shear stress and its nondimensionalized value
  are $\Ev{\sigma_{xz}}{\ze=0} = \rhoGxH$
  and $\Ev{\tilde\sigma}{\ze=0} = \rhoGxH/S$.
For the paste to flow steadily,
  this wall shear stress $\rhoGxH$ must be greater
  than the yield stress $\Yield$.
The maximum $\alpha$
  also occurs at the wall:
\begin{equation}
 \max_\ze \alpha 
  = -{\frac12} 
  \log\left[\,
       1 - \left( \frac{\rhoGxH}{S} \right)^2
       \;\right]
  \label{z4}     
\end{equation}
  according to Eq.~(\ref{z3}).

The above discussion suggests
  two nondimensional parameters 
  that can be expressed as a ratio $\rhoGxH : \Yield : S$.
Let us complete the dimensional analysis of Eqs.~(\ref{eqs:1D})
  before proceeding to the numerical calculation
  of time-dependent solutions.

\subsection{Dimensional analysis}
\label{subsec:dim}

Eqs.~(\ref{eqs:1D}) contain five physical parameters,
  namely $S$, $\etaP$, $\rho$, $G_x$ and $\Yield$ (the
  last one comes through $\nu$).
The first three determine
  the viscoelastic time scale $\Tau = \etaP/S$
  and the length scale $\Ell = \etaP/\sqrt{{\rho}{S}}$.
The boundary conditions introduce
  the layer thickness $H$ as another length scale.

The system is characterized by three nondimensional parameters,
  for example, $H/\Ell$, $\Yield/S$, and $\rhoGxH/S$ (or
  a suitable combination of them).
Note that $\rhoGxH$ gives an estimation of the wall shear stress,
  which represents the magnitude of the external forcing.

Evaluation of Reynolds number will be useful
  for considering the Newtonian limit.
On the basis of $\etaP$, $H$, and 
  $U \sim \rhoGxH/(\etaP/H) = \rho G_x H^2 / \etaP $,
  it is estimated as 
\[
  R_H \sim \left( \frac{H}{\Ell} \right)^2 \frac{\rhoGxH}{S};
\]
  this is indeed calculated
  from two of the three parameters stated above.
In the present setup, $H$ is taken 
  as the representative length scale,
  but we point out a general possibility 
  that the system may be characterized by other Reynolds numbers,
  such as $R_L = UL/\etaP$
  based on the horizontal length scale $L$.
In future studies
  this point may have to be taken into account.


\begin{figure*}
 \includegraphics[clip,width=13.5cm]{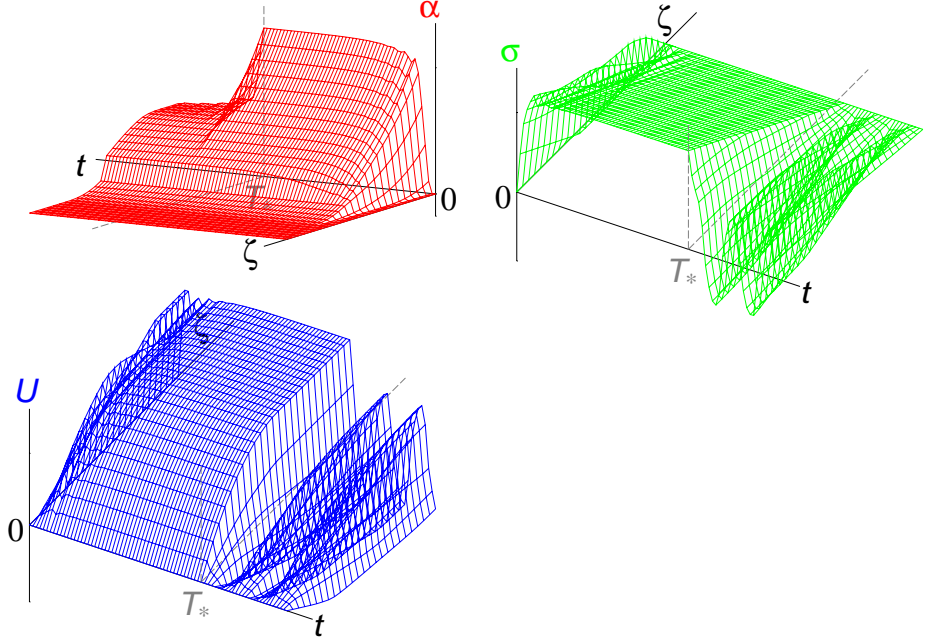}  
 \caption{Typical evolution of $(\alpha, \tilde\sigma, U)$.}%
 \label{Fig:E}%
\end{figure*}

\begin{figure}
 \begin{center}
  \includegraphics[clip,width=7.0cm]{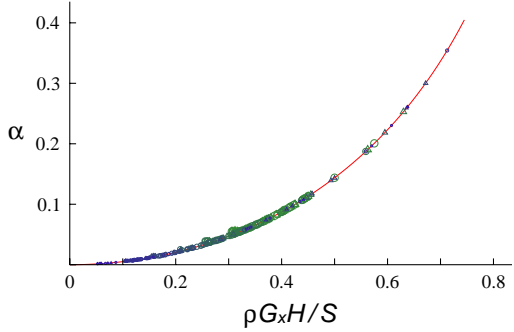}  
  \caption{%
  Steady values of $\alpha$ during the forcing.
  Spatial maximum of $\alpha$,
  whose steadiness is checked for a certain time range,
  is plotted against $\rhoGxH/S$.
  The line shows the analytical solution~(\ref{z4}).
  The circles and the triangles represent
  numerical values for $T_* = 100\Tau$ and $T_* = 200\Tau$,
  respectively.
  The size and the color of the symbols indicate $\Yield/S$,
  from the small blue symbols for $\Yield = 0.05\,S $
  to the large green symbols for $\Yield = 0.30\,S $.
  }
  \label{Fig:A1}
 \end{center}
\end{figure}

\begin{figure}
 \includegraphics[clip,width=7.0cm]{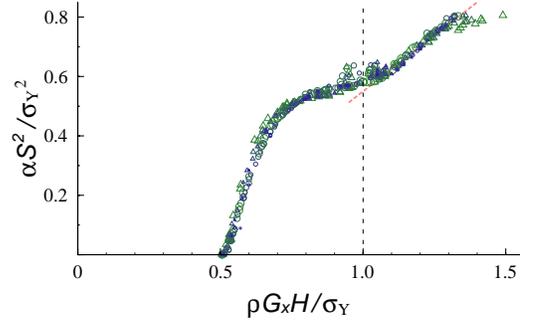}  
 \caption{\label{Fig:A2}%
 Residual value of $\alpha$ that remains in the paste
 after the forcing is removed.
 The spatial maximum of $\alpha$, rescaled by $(\Yield/S)^2$,
 is plotted against $\rhoGxH/\Yield$.
 The same symbols (circles and triangles) are used
 as in Fig.~\ref{Fig:A1}.
 The nonvertical broken line, with slope $0.75$,
 represents the fitting relation~(\ref{*r}).
 }%
\end{figure}


\subsection{Numerical calculation of unsteady solution}

The author calculated
  the numerical solutions of Eqs.~(\ref{eqs:1D}) (slightly modified,
  as we will see below)
  under the initial condition
  $\Ev{(\alpha,\tilde\sigma,U)}{t=0} = (0,0,0)$
  and the boundary conditions (\ref{BC1}) and (\ref{BC2}),
  with $\nu(\Eps(\alpha,\tilde\sigma))$
  defined by Eqs.~(\ref{*3}) and (\ref{Eps}),
  for hundreds of different nondimensional parameters.
With the hyperbolic character of Eqs.~(\ref{eqs:1D}) 
  taken into account,
  the calculation adopted 
  the two-step Lax-Wendroff scheme~\cite{NumericalRecipe.Book1988}.

Since we are interested not only in the creation process of $\alpha$,
  but also the storage of $\alpha$ after the flow is stopped,
  it is necessary to simulate
  the process to stop the flow.
To this aim, we ``switch off'' gravity
  at some time $t = T_* \; (\gg \Tau)$,
  replacing Eq.~(\ref{*u}) by 
\begin{equation}
  \dt{U}
  = {\frac{S}{\rho}} \Dz{\tilde\sigma} + \frac{F_x}{\rho},
  \quad
  F_x = \begin{cases}
          \rho G_x  &  (0 < t < T_*)  \\
          0         &  (t > T_*)
        \end{cases}
\end{equation}
  with $T_* = 100\Tau$ or $T_* = 200\Tau$.

Fig.~\ref{Fig:E} depicts
  a typical evolution of $(\alpha,\tilde\sigma,U)$.
The parameters are
  $ H = 5\Ell$, \  
  $\rhoGxH:\Yield:S = {0.6} : {0.3} : 1$, \ 
  and $T_* = 100\Tau$.
In the first stage of the evolution,
  the system rapidly approaches steady state,
  except for the region adjacent to the boundary between the
  fluidized and solidified regions ($\ze = \zYield = 2.58\,\Ell$)
  where the relaxation time is significantly longer.
After the gravity is ``switched off'' at $t = T_*$,
  both $\tilde\sigma$ and $U$ oscillates around zero.
This oscillation should be damped 
  if we consider the solvent viscosity,
  which is neglected in the present model.
What must be noted is that $\alpha$ remains finite,
  though it decreases,
  after the driving force is switched off at $t = T_*$.
The sign of the residual $\alpha$ is positive.

According to the analysis
  of 801 cases with $T = 100\Tau$
  and 689 cases with $T = 200\Tau$,  
  the behavior of $\alpha$ for different values of $\rhoGxH$
  is summarized as follows.
For $\rhoGxH$ smaller than $\Yield/2$,
  throughout the evolution $\alpha$ remains zero. 
If $\rhoGxH$ exceeds $\Yield/2$ but still remains below $\Yield$,
  the evolution during the forcing ($0<t<T_*$) 
  is basically unsteady, 
  where $\alpha$ is produced little by little
  from the interference of the stress waves.
For $\Yield < \rhoGxH < S$ (we always assume $\Yield < S$),
  steady yield flow occurs,
  creating $\alpha$ according to Eq.~(\ref{z3}).
In both regimes stated above,
 a residual $\alpha$ is observed after $t=T_*$.
The steady solution, Eq.~(\ref{z3}),
  ceases to exist for $\rhoGxH > S$,
  which leads to the unlimited acceleration of the flow.
This last case is out of the scope of the present model,
  because $U$ should be limited
  if, again, the solvent viscosity is taken into account.

Steady solutions obtained by the time-dependent calculation
  during the forcing, approximately for $T_* / 2 < t < (3/4)T_*$,
  are checked against the analytical solution
  in Fig.~\ref{Fig:A1}.
The curve shows
  $\max_\ze {\alpha}$ given by Eq.~(\ref{z4})
  as a function of $\rhoGxH/S$.
The symbols,
  consisting of 122 circles ($T_* = 100\Tau$) 
  and 110 triangles ($T_* = 200\Tau$),
  indicate the numerical values of $\max_\ze {\alpha}$
  calculated within the range $\Yield < \rhoGxH < 0.72\,S $, 
  $0.05\,S \le \Yield \le 0.30\,S$,  
  and $0.20\,\Ell \le H \le 8.0\,\Ell$.
The size (and the color) of each symbol
  indicates the magnitude of $\Yield/S$.
Fig.~\ref{Fig:A1} demonstrates
  that the value of steady $\alpha$ is independent of $\Yield$,
  once $\rhoGxH$ exceeds it.
For $T_* = 100\Tau$,
  there were several cases for which $\alpha$ did not attain
  its steady value (with the criterion
  $\Delta{\alpha}/|\alpha| = 0.02$);
  these cases are eliminated from Fig.~\ref{Fig:A1} for clarity.
Such exceptional cases did not occur
  for $T_* = 200\Tau$.

As an explanation of the Nakahara effect,
  it is essential to show that some of $\alpha$ remains in the paste 
  even after the flow is stopped,
  instead of decaying away.
Fig.~\ref{Fig:A2} shows the numerical values of $\max_\ze {\alpha}$
  remaining steady (not to decay any more) after the flow is stopped.
Here not $\alpha$ itself, but $\alpha S^2/\Yield^2$
  is plotted against $\rhoGxH/\Yield$
  for $\rhoGxH > \Yield/2$ (the ranges of $\Yield/S$ is 
  the same as in Fig.~\ref{Fig:A1},
  and that of $H/\Ell$ is $0.05\,\Ell \le H \le 8.0\,\Ell$).
The values of residual $\alpha$ for $\Yield < \rhoGxH < 1.3\,\Yield$
  is fitted by
\begin{equation}
  \max_\ze {\alpha} \sim 
  \frac{{\Yield}^2}{S^2} 
  \left( 0.75\,\frac{\rhoGxH}{\Yield} - 0.2 \right)
  \label{*r}.
\end{equation}
In contrast to the steady value of $\alpha$
  in the flow subject to the driving force,
  the residual value of $\alpha$ in Eq.~(\ref{*r})
  is strongly dependent on $\Yield$.
In particular, if $\rhoGxH/\Yield$ is kept constant,
  Eq.~(\ref{*r}) states that the residual value of $\alpha$
  is scaled by $(\Yield/S)^2$.
This result seems understandable
  if we assume that,
  during the decay of $\alpha$ and $\tilde\sigma = {\sigma_{xz}}/S$,
  the first equal sign in Eq.~(\ref{z3}) remains valid,
  until $\sigma_{xz}$ reaches the threshold value $\Yield$.
This gives a rough estimation of
  the residual $\alpha \simeq 0.5\,(\Yield/S)^2$.
Unfortunately, 
  theoretical clarification of Eq.~(\ref{*r})
  in regard to its dependence on $\rhoGxH/\Yield$
  is not currently available.


\section{Discussion and concluding remarks}

\subsection{Relationship with crack pattern experiments}
\label{subsec:crack}

In this paper we have found the
  creation and fixation of the $x$-directional tension
  using a model equation for flows of isotropic pastes. 
This provides a possible scenario 
  for the Nakahara effect (Type I).

During the drying process,
  the paste slowly shrinks.
Mathematically, this process is described
  as an isotropic contraction (shrinking) 
  of the natural metric $\mathsf{g}^\Nat$.
If the paste had not undergone a flowing process,
  this contraction would produce 
  a basically isotropic tension
  in the $(x,y)$-plane (parallel to the surface and the bottom) 
  and therefore would lead to isotropic crack patterns.
Actually, this is not the case:
  we have found that 
  a positive $\alpha$ is created during the flowing process,
  which implies that the natural metric 
  is already \emph{contracted} in the streamwise direction.
Strictly speaking,
  the present analysis is limited
  to the two-dimensional system in the $(x,z)$-plane
  and therefore it cannot tell
  whether any $y$-directional contraction occurs,
  but it is unlikely that it will occur 
  to the same extent as the $x$-directional contraction.
In fact, though a full analysis of three-dimensional system
  is too complicated to develop here,
  a simple perturbation analysis supports the above conjecture.
The bonds perpendicular to the flow
  are therefore the first ones to break,
  causing cracks perpendicular to the flow
  and thus clarifying the Nakahara effect.

The present numerical analysis predicts 
  that the magnitude of the residual $\alpha$
  is scaled by $(\Yield/S)^2$,
  as is seen in Fig.~\ref{Fig:A2} and Eq.~(\ref{*r}).
This result is consistent 
  with the observation of Nakahara and Matsuo
  in regard to the strength of the memory effect
  summarized as Fig.~2 in Ref.~\cite{Nakahara.JPSJ74}.
The figure presents 
  a classification of the observed patterns
  as a function of the solid volume fraction (density of the paste)
  and the strength of the external forcing.
Its Region B, which lies just above the yield stress line
  and exhibits the memory effect,
  is subdivided
  according to the strength of the anisotropy in the pattern;
  strong anisotropy is observed
  for denser pastes (lamellar crack patterns, denoted by solid squares,
  occupy the subregion with volume fraction greater than 40\%),
  while less dense pastes exhibit weaker anisotropy,
  resulting in large-scale lamellar cracks ($\gg H$)
  combined with cellular structure
  with smaller length scales ($\sim H$).
If we admit that $\Yield/S$ is greater for denser pastes,
  the difference in the strength of anisotropy 
  can be explained from our theory
  predicting $\alpha \propto (\Yield/S)^2$.

\subsection{Comparison with dry granular flows and other systems
            exhibiting memory effects}
\label{subsec:granular}

Memory effects are quite common in many glassy systems,
  ranging from granular matters to spin glasses.
In the case of dry granular matters~\cite{Duran.Book2000,Aranson.RMP78},
  history-dependent behavior essentially
  originates from the existence
  of interparticulate static friction.
According to Coulomb's friction law,
  the interparticulate forces admit static indeterminacy,
  giving rise to the history-dependent stress state.
Fluidization and solidification of granular matter 
  also involves the creation and destruction of grain-scale structures,
  such as arching and force chains.
Though the present study on paste flows 
  is based on macroscopic description
  and therefore
  discussion on the grain-scale structure is outside its scope,
  comparative consideration on static indeterminacy
  is quite helpful in understanding 
  some common mechanisms underlying paste flows and dry granular flows.

As seen at the top of Sec.~\ref{sec:analysis},
 the threshold mechanism in $\nu(\Eps)$
  results in the static indeterminacy of $\alpha$.
It is this static indeterminacy
  that enables the retention of the memory of the shear flow.
(In a different setup~\cite{Ooshida.PRL95},
  residual stress is introduced via static indeterminacy of $\beta$.)
Thus the present paste model shares an important feature
  with dry granular systems.

\begin{figure}
 \includegraphics[clip,width=7.5cm]{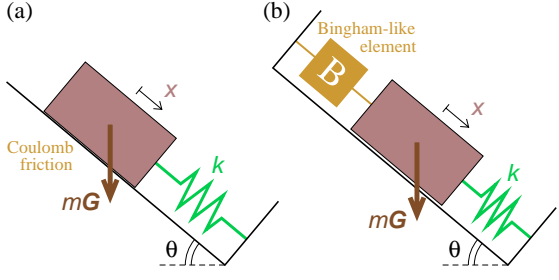}
 \caption{\label{Fig:D}%
 Duran's brick on a slope to illustrate static indeterminacy.
 (a) Original setup due to Duran~\cite{Duran.Book2000}.
     Besides the gravity and the repulsion of the spring,
     the brick is subject to static Coulomb friction.
 (b) A modified setup, where Coulomb friction is replaced
     by a Bingham-like elastoplastic element. 
 }%
\end{figure}

To elucidate the analogy and distinction
  between the Bingham plasticity and Coulomb friction,
  let us consider an instructive problem 
  taken from Chapter 3 of Duran's book~\cite{Duran.Book2000}.
Suppose a brick on an inclined wall,
  subject to static Coulomb friction (coefficient $\muS$) and a spring,
  as is illustrated in Fig.~\ref{Fig:D}(a).
Duran's problem is to determine the deformation $x$ (or equivalently
  the repulsion $kx$) of the spring
  as a function of the inclination angle $\theta$,
  when $\theta$ varies slowly in time.

\begin{figure}
 \includegraphics[clip,width=8.0cm]{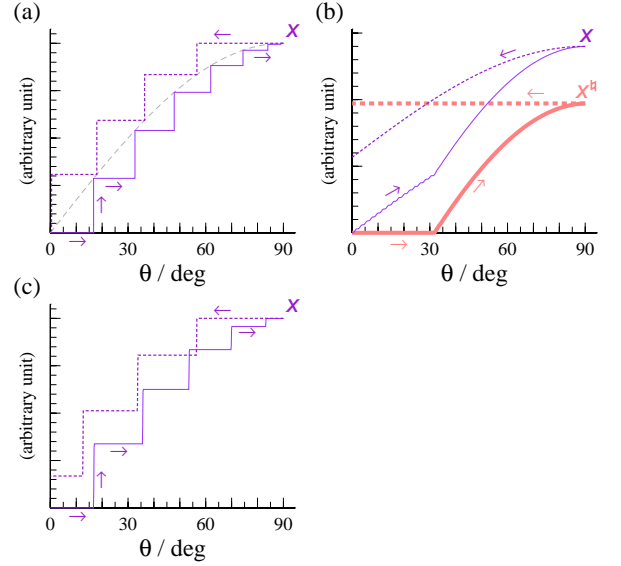}
 \caption{\label{Fig:H}%
 Mechanical hysteresis in Duran's brick.
 The deformation $x$ is plotted against the inclination $\theta$,
 with a solid line for the forward process (increasing $\theta$) 
 and with a broken line for the backward process.
 (a) Duran's solution~\cite{Duran.Book2000}
 in the case of static Coulomb friction.
 (b) Bingham-like case with Eqs.~(\ref{eqs:Duran+Bingham}).
 In addition to $x$,
 the natural length of the Bingham-like element, $\xNat$,
 is delineated in thick (red) lines.
 (c) A solution of Eq.~(\ref{h4})
 exhibiting Coulomb-like stick-slip motion.
 }%
\end{figure}

Suppose
  that the wall starts from the horizontal position ($\theta=0$)
  and that we know the initial value of $x$, which we denote by $x_0$.
For a while $x$ is stuck to $x_0$,
  until the ``yield'' criterion
\[
  \left| mG\sin\theta - kx \right| = \muS mG\cos\theta
\]
  is attained and the brick starts to slip.
We assume the viscous resistance $-c\dot{x}$
  and neglect the dynamic Coulomb friction for simplicity~\footnote{%
    The presence of the viscous drag is not explicitly stated
    in Duran's book~\cite{Duran.Book2000},
    but it seems to be implicitly assumed
    by stating that $x$ stops at the position
    satisfying $kx = mG\sin\theta$. 
  },
  so that the brick moves according to 
\[
  m\frac{\D^2{x}}{{\D{t}}^2} = -c\DD{x}{t} - kx + mG\sin\theta
  \quad (\text{brick in motion})
\]
  and eventually stops.
This process is repeated while $\theta$ increases, 
  as is shown in Fig.~\ref{Fig:H}(a) with a solid line (each slip
  is assumed to stop when $kx = mG\sin\theta$
  according to Duran~\cite{Duran.Book2000}).
If $\theta$ starts from $\pi/2$ and decreases slowly in time,
  a similar but different stick-slip motion occurs,
  as depicted by the broken line.
Thus, the system exhibits
  mechanical hysteresis due to static friction.

Now let us compare this mechanical hysteresis
  with the behavior of the system in Fig.~\ref{Fig:D}(b),
  where the Coulomb friction is replaced by
  a discrete-element analogue of Bingham-like elastoplasticity.
Its behavior 
  is defined by combining Eqs.~(\ref{d4}) and (\ref{eqs:d5}) with
\begin{equation}
  \tau^{-1} = \nu(T)
  \sim \begin{cases}
               0           & (\, |E(T)| <   \text{threshold} ) \\
               \Tau^{-1}   & (\, |E(T)| \gg \text{threshold} ) 
       \end{cases}
  \label{h1}
\end{equation}
  which is a discrete-element version of Eq.~(\ref{*1}),
  with $E(T) = T^2/(2\kappa)$
  standing for the elastic energy stored in this element.
The governing equation of this system
  is summarized as
\begin{subequations}%
 \begin{gather}
   \DD{\xNat}{t} = \nu(T) \left( x - \xNat \right)     \label{h2}, \\ 
   m\frac{\D^2{x}}{{\D{t}}^2} = -T -kx + mG\sin\theta  \label{h3},
 \end{gather}%
 \label{eqs:Duran+Bingham}%
\end{subequations}%
  supplemented with $T = \kappa\,(x-\xNat)$.  
As for $\nu(T)$,
  a Lipschitz-continuous form analogous to Eq.~(\ref{*3}) is assumed.
By numerical integration of Eqs.~(\ref{eqs:Duran+Bingham})
  with $\theta$ increased slowly from zero to $\pi/2$
  and then decreased back,
  we obtain the result shown in Fig.~\ref{Fig:H}(b).
The thick solid line indicates
  that a shift of $\xNat$ has occurred 
  during the process of increasing $\theta$, 
  and this shift was not recovered at all  
  when $\theta$ was decreased (thick broken line).
In addition, when $\theta$ has returned to zero,
  there remains a difference in $x$ and $\xNat$,
  indicating residual pressure in this case.
Thus, again, a hysteresis due to static indeterminacy
  is observed.
There is an important difference, however,
  that the curves in Fig.~\ref{Fig:H}(b)
  are much less singular than those in Fig.~\ref{Fig:H}(a).
In other words,
  at least for the values of the parameters
  and the functional form of $\nu(T)$ used in this calculation,
  no stick-slip behavior is observed.
This is probably related to the property of the Bingham model, 
  which predicts continuous shear stress across the yield front
  in quite general cases~\cite{Sekimoto.JNNFM46}.

It is an interesting attempt 
  to reformulate
  the stick-slip motion subject to static Coulomb friction
  in terms of $\nu$,
  to obtain a (formally) unified equation:
\begin{equation}
  \DD{v}{t} = \nu(v,F,N) \left( \frac{F}{c} - v \right)   \label{h4}
\end{equation}
  with 
\[
  v = \DD{x}{t},          \quad
  F = -kx + mG\sin\theta, \quad
  N =       mG\cos\theta.
\]
A naive choice for $\nu$ is
\[
  \nu(v,F,N) = \begin{cases}
                0                &  ({v=0}\ \text{and}\ {|F| < {\muS}N})\\
                \Tau^{-1} = c/m  &  (\text{otherwise}).
               \end{cases}
\]
Though this function is too singular
  to constitute a mathematically sound evolutional equation,
  adoption of continuous interpolation similar to Eq.~(\ref{*3})
  enables the numerical integration of Eq.~(\ref{h4}),
  resulting in stick-slip motion shown in Fig.~\ref{Fig:H}(c).
Note that, in Eq.~(\ref{h4}),
  the relaxation is attributed to the momentum,
  but this seems to be somewhat unnatural
  if we consider that friction is a property of the interface
  while momentum concerns the whole mass of the body.
Rather,
  in analogy to Eqs.~(\ref{eqs:Duran+Bingham})
  where relaxation is attributed to $\xNat$,
  it seems more appropriate to introduce
  a variable describing the state of the interface (possibly
  similar to the one 
  introduced by Carlson and Batista~\cite{Carlson.PRE53})
  and prescribe its relaxation.
This is 
  beyond the scope of the present work, however.

In soil mechanics,
  a continuum version of Coulomb friction 
  is known as Mohr-Coulomb plasticity~\cite{Nedderman.Book1992}.
Its application to the statics of granular materials
  is usually supplemented with the limit-state assumption,
  which states
  that the ratio of the shear stress to the normal stress
  is just below the threshold value everywhere.
This assumption makes it possible
  to evaluate the stress field
  without introducing granular elasticity 
  that is not understood very well.
However, 
  this theory encounters a number of difficulties,
  as is discussed by Kamrim and Bazant~\cite{Kamrin.PRE75}.
According to this theory, 
  the static stress field
  is subject to a nonlinear hyperbolic system of equations 
  (not in space-time but in the $(x,y)$-plane),
  which predicts a highly discontinuous stress field.
The solution for the velocity field can be even more singular,
  which seems abnormal both physically and mathematically.
Sometimes it also fails to satisfy
  the boundary conditions.
Kamrim and Bazant~\cite{Kamrin.PRE75} have shown
  that these difficulties can be avoided,
  within the framework of Mohr-Coulomb plasticity
  with the limit-state assumption,
  by introducing diffusive motions via mesoscale objects called ``spots.''
In spite of this successful result,
  the theory is not free from the limitation
  due to the limit-state assumption,
  as the authors themselves admit
  that clearly it breaks down in some cases. 

Kamrim and Bazant~\cite{Kamrin.PRE75} state repeatedly
  that the introduction of elasticity will solve
  the difficulties of the Mohr-Coulomb plastic model. 
To some extent,
  this remark applies to Bingham plasticity as well.
For example,
  the original Bingham model 
  exhibits a singular behavior due to the lack of elasticity,
  in the sense that the propagation speed of yield front 
  is infinitely large~\cite{Sekimoto.JNNFM39}.
Treatment of residual stress
  would be also very difficult, if not impossible,
  without considering finite elasticity.
This is why we sought to develop an elastoplastic paste model 
  from the beginning.

The present theory 
  is conceptually akin to the models of the memory effect
  in polymeric materials~\cite{Miyamoto.PRL88,Ohzono.PRE72}.
Miyamoto \textit{et al.}~\cite{Miyamoto.PRL88} studied 
  the memory effect in the glass transition of vulcanized rubber.
They explained their experimental results
  with a Maxwell-like model,
\begin{align}
 \sigma(t)
 &= \sigma_{\text{rubber}}(\Temperature(t),\gamma(t))  
 \notag \\ & \qquad 
 + S_{\text{glass}} 
 \int_{-\infty}^{t}
 [\gamma(t)-\gamma(t')] \dd{\KernelG{\tilde{t}}}{t'}\,\D{t'}
 \label{Miyamoto},
\end{align}
  where $\sigma$ is stress,
  $\Temperature$ is temperature,
  $\gamma$ is strain, 
  $\KernelG{\;\cdot\;}$ is normalized relaxation function,
  and $\tilde{t}$, defined by
\[
  \tilde{t} = \tilde{t}(t,t') 
  = \int_{t'}^{t} \frac{\D{u}}{\tau(\Temperature(u),\gamma(u))},
\]
  stands for the intrinsic time lapse.
The effect of temperature control (quenching and reheating)
  is expressed via $\tau$,
  which changes the pace of the intrinsic time $t'$
  and thereby affects the memory function in Eq.~(\ref{Miyamoto}).
Note that $\gamma(t')$ in the integral
  can be read as the natural length of a spring born at the time $t'$.
In this sense, Eq.~(\ref{d4}) can be regarded
  as a simplified version of Eq.~(\ref{Miyamoto}),
  though there is an important difference
  that the memory in Eq.~(\ref{d4}) 
  is ascribed to a single variable $\xNat$,
  while Eq.~(\ref{Miyamoto}) can memorize
  more about the history of $\gamma(t')$.
The ``memory capacity'' of Eq.~(\ref{Miyamoto})
  depends on the property of
  the relaxation function $\KernelG{\;\cdot\;}$.
Using a sum of two exponential functions,
  which implies double relaxation,
  Miyamoto \textit{et al.}~\cite{Miyamoto.PRL88}
  has successfully reproduced the memory effect,
  including the effect of aging.
In the present model, contrastively,
  the relaxation time $\tau$ is assumed to be a single scalar function.
Instead, the spatial distribution of $\alpha$
  and the effects of nonlinear elasticity are taken into account,
  thus enabling the creation and storage of the streamwise tension.

The idea of ascribing the memory
  to the plastic shift in the neutral point of elasticity,
  corresponding to $\alpha$ in Eqs.~(\ref{*P}) and (\ref{*a}),
  $\xNat$ in Eqs.~(\ref{d4}) and (\ref{h2}), 
  and $\gamma(t')$ in Eq.~(\ref{Miyamoto}),
  is also shared by Ohzono \textit{et al.}~\cite{Ohzono.PRE72}.
They studied microwrinkle patterns
  produced on a platinum-coated elastomer surface,
  governed by the competition
  between the restoring force of the platinum tending to be less curved
  and that of the elastomer that aims to shrink back.
At room temperature,
  application of a uniaxial compression force
  breaks the force balance and changes the wrinkle pattern,
  but the original pattern is retrieved 
  after the external force is removed. 
Contrastively, a protocol involving higher temperature
  (annealing-cooling-unloading protocol)
  changes the wrinkle pattern, introducing strong anisotropy.
The new pattern is less stable to external forcing at room temperature,
  suggesting the presence of multiple metastable states.
These experimental results are compared with a model
  prescribing the minimization of the elastic energy
  as a functional of the surface elevation $z = z(x,y)$,
\begin{gather}
  U[z] 
  = U_{\text{bending}} 
  + U_{\text{in-plane}} + U_{\text{substrate}} \notag, \\ 
  U_{\text{substrate}} 
  = \int \left\{ a\,(z-\zm)^2 + b\,(z-\zm)^4 \right\} \,\D{x}\D{y}
  \label{Ohzono},
\end{gather}
  where 
  $U_{\text{bending}}$ and $U_{\text{in-plane}}$
  are the potentials of bending and in-plane deformation 
  of the platinum layer,
  $U_{\text{substrate}}$ is the potential of the substrate (with
  the constants $a$ and $b$ specified explicitly
  in terms of the material constants of the elastomer),
  and $\zm = \zm(x,y)$ represents
  the neutral point of $U_{\text{substrate}}$.
Correspondence with Eq.~(\ref{d4})
  is obvious.
The memory is carried by spatial distribution of $\zm$,
  which is fixed at the room temperature
  but is subject to plastic flow 
  in the annealing-cooling-unloading protocol.

Generally,
  in elastic systems with more than several degrees of freedom (and
  particularly in continua),
  a shift in the neutral point introduces mechanical frustration.
In the case of Ohzono \textit{et al.}~\cite{Ohzono.PRE72},
  it modifies the existing frustration, introducing multiple stability.
In addition, 
  spatial heterogeneity of $\alpha$ and $\beta$ in Eq.~(\ref{g1})
  is equivalent to the continuous distribution
  of edge dislocations and screw dislocations,
  respectively~\cite{Ooshida.PRL95,Landau.elasticity,Chaikin.Book1995}.
Thus, frustration is observed universally
  in systems admitting plasticity (in any sense of the word),
  ranging from granular matters to metal crystals and spin glasses.
From this viewpoint, we understand Eq.~(\ref{*a}) 
  as describing the dynamic creation and static retention 
  of mechanical frustration,
  presenting a macroscopic analogue of dislocation dynamics.


\subsection{Future directions}
\label{subsec:future}

The present study
  is entirely based on macroscopic phenomenology.
It predicts the \emph{presence} of a macroscopic mechanism
  that leads to the Type-I Nakahara effect,
  but it does not assert the \emph{absence} of other mechanisms,
  such as the creation of bond fabric or microscopic texture.
A possible scenario
  is that the microscopic bond structure is well represented
  by the macroscopic (hydrodynamic) variables,
  such as $\alpha$ and $\tilde{\sigma}$,
  so that the most important feature of the mechanism
  is already captured by the hydrodynamic equations.
In other words,
  we expect something analogous to ferromagnetism,
  where the macroscopic magnetization represents the order parameter.
We cannot deny the possibility
  that some pastes have ``anti-ferromagnetic'' bond structure,
  which makes the macroscopic description more difficult.
Even in the ``ferromagnetic'' case,
  consideration of microscopic details may introduce 
  some modification.
For example, we have regarded the yield stress as a given constant,
  but this may possibly need to be modified,
  in the way similar to work hardening and the Bauschinger effect
  in metals~\cite{McLean.Book1962}.
The constitutive relations assumed in this paper
  require justification more certain than a physicist's intuition,
  either by microscopic analysis or thermodynamical inspection.
It is worthwhile to consider
  an extension of microscopic theories for glassy liquids,
  such as the mode-coupling theory~\cite{Miyazaki.PRE66,Fuchs.PRL89}
  and the pair distribution function theory~\cite{Otsuki.JStat2006},
  in the direction corresponding
  to that of direct-interaction approximation in fluid turbulence
  based on Lagrangian description~\cite{Frisch.Book1995,%
  Kraichnan.PhF8,Kaneda.JFM107,Kida.JFM345},
  in search of microscopic expression for $g_\Nat^{ij}$.
Such a microscopic approach would also allow us
  to construct a model for pastes whose properties are not isotropic.
It is expected, for example,
  that a model including
  competitive interaction between the natural metric and director field
  may clarify the Type-II Nakahara effect.

Within the framework of the present model,
  an explanation of Eq.~(\ref{*r})
  that gives the magnitude of the residual $\alpha$
  is an open question.
It is also necessary
  to extend the present work in several respects.
On one hand,
  the limitation to uniform flows must be removed.
The stopping process is simulated in this paper
  by switching gravity off,
  but in real experiments 
  the paste flow stops when the paste supply is cut.
Simulation of this process requires
  the introduction of a variable layer thickness $h=h(x,t)$,
  where the flow and the stress fields depend on $x$ as well.
This extension will clarify 
  the relevance of different mechanisms,
  such as the one proposed by Otsuki~\cite{Otsuki.PRE72}
  where the $x$-dependence of the plastic deformation is essential.
The present model
  can be readily extended in this direction,
  though its numerical analysis will be much more difficult.
Derivation of reduced equations,
  such as depth-averaging (corresponding to Shkadov model~%
  \cite{Shkadov.IANMZhG1,Ruyer-Quil.EurPhysJB6,Chang.Book2002} 
  in the film flows 
  and Saint-Venant model~\cite{Saint-Venant.Paris1871,Forterre.JFM486} 
  in civil engineering),
  will be worth considering.

On the side of experiments,
  it is desirable to realize a uniform slope flow
  by eliminating the boundary effect in the $y$ direction.
It is also necessary to measure the paste properties,
  such as $S$ and $\Yield$,
  so that a qualitative comparison 
  between the theory and the experiment becomes possible.
Finally, since the mechanism proposed in this paper
  is closely related to the nonlinear viscoelasticity,
  it will be highly supportive
  to detect any indication of nonlinear viscoelasticity
  in the paste, such as Weissenberg effect.


\begin{acknowledgments}
 The author expresses his cordial thanks
   to Sin-ichi Sasa and Ken Sekimoto
   for their helpful comments and encouragement.
 The author is also grateful 
   to Akio Nakahara, Michio Otsuki,
   Takahiro Hatano, Shio Inagaki,
   Takeshi Matsumoto, Yasuhide Fukumoto,
   Christian Ruyer-Quil, Takuya Ohzono,
   Hiizu Nakanishi, Hisao Hayakawa,
   Yasuhiro Oda,
   Hiromitsu Kawazoe, and Kenta Kanemura 
   for their insightful comments and discussions,
   and in particular to Motozo Hayakawa 
   for providing the author with several useful comments,
   including Ref.~\cite{McLean.Book1962}.
 This work was supported
   by a Grant-in-Aid for Young Scientists (B),
   No.~18740233, MEXT (Japan).
\end{acknowledgments}

\appendix

\section{Notes on the formulation of Lagrangian continuum mechanics
         in terms of differential geometry}
\label{app:diff}

Here we summarize the minimal mathematical knowledge
  required to understand,
  for example, how to calculate each side of Eq.~(\ref{d1}).
Instead of going along the rather expensive highway
  of Riemannian differential geometry,
  we take a shortcut,
  making full use of the $\nd$-dimensional Euclidean space
  where the whole system is embedded.

\subsection*{Contravariant vector components}

For each instant (with $t$ fixed arbitrarily),
  the mapping from $\BoldXi$ to $\rvect$
  provides an instantaneous curvilinear coordinate system.
This is sometimes refered to
  as a \emph{convected coordinate system}~\cite{Bird.Book1987}.
It is this coordinated system, and not the space, 
  that is curved.

Provided that the mapping (\ref{k1}), for fixed $t$,
  is sufficiently smooth and locally invertible,
  we find that
\[
  \left\{ 
    \dd{\rvect}{\xi}, \dd{\rvect}{\eta}, \dd{\rvect}{\ze} 
  \right\}
  \quad
  (\mbox{for $\nd=3$})
\]
  forms a set of local bases
  in the $\nd$-dimensional Euclidean space ($\rvect$-space).
Then, an arbitrary vector field, say $\mb{f}$,
  can be expressed as
\[
  \mb{f} 
  = \begin{bmatrix}
     \displaystyle\dd{\rvect}{\xi}  & 
     \displaystyle\dd{\rvect}{\eta} & 
     \displaystyle\dd{\rvect}{\ze}
    \end{bmatrix}
    \begin{bmatrix} f^\xi \\ f^\eta \\ f^\ze \end{bmatrix} 
  = f^\xi  \dd{\rvect}{\xi}
  + f^\eta \dd{\rvect}{\eta}
  + f^\ze  \dd{\rvect}{\ze}
\]
  or, in abbreviation with Einstein's contraction rule,
\begin{equation}
  \mb{f} = f^i \B{i}   \label{contravariant}.
\end{equation}
The coefficients $(f^i)$ in Eq.~(\ref{contravariant})
  are referred to as \emph{contravariant components}
  of the vector field $\mb{f}$.
According to the convention of differential geometry,
  the contravariant components are superscripted.

If the labeling variable
  is changed from $\BoldXi$ to $\bar{\BoldXi}$ (in terms
  of a continuous, one-to-one mapping independent of $t$),
  the bases are changed to
\[
  \dd{\rvect}{\bar\xi^i} 
  =
  \dd{\xi^j}{\bar\xi^i}\dd{\rvect}{\xi^j}.
\]
Meanwhile the change from $(f^i)$ to $(\bar{f}^i)$
  occurs in such a way that
  it cancels the change in the bases (therefore
  the name ``contravariant''),
  so that the vector $\mb{f}$ itself remains unaffected:
\[
  \mb{f} = f^i       \dd{\rvect}{\xi^i}  
         = \bar{f}^i \dd{\rvect}{\bar\xi^i}. 
\]
An equation describing the relations between physical quantities
  should be independent of the choice of labeling variables.
This is assured 
  if and only if every term on both sides of the equation 
  has the same behavior in regard to the relabeling.
For example,
\[
  a^i = 2 b^i
\]
  is acceptable, while
\[
  a^i \stackrel?= b^i + 2
\]
  is not (we cannot add a scalar $2$ 
  to a contravariant vector component $b^i$).

A second-order tensor, say $\TensorP$,
  can be expressed as 
\begin{equation}
  \TensorP = P^{ij} (\B{i})\otimes(\B{j})
\end{equation}
  where $\otimes$ is the tensor product,
  such that
\[
  \mb{a}\cdot(\mb{b}\otimes\mb{c})
  = (\mb{c}\otimes\mb{b})\cdot\mb{a}
  = (\mb{a}\cdot\mb{b})\,\mb{c}.
\]
The contravariant components $(P^{ij})$
  are subject to the same kind of change
  as the product of two contravariant vector components,
  so that $\TensorP$ remains unaffected by the relabeling.
Note that Kronecker's delta with superscripts,
  $\delta^{ij}$,
  does not behave properly in regard to relabeling
  and therefore is not acceptable as a physically meaningful tensor.

\subsection*{Dual basis and covariant vector components}

As has been stated,
  we assume that the mapping from $\BoldXi$ to $\rvect$ 
  is smooth and invertible.
Therefore, it makes sense to define
\begin{equation}
  \nabla\xi^i = \dd{\xi^i}{\rvect};
\end{equation}
  a nabla without subscript, $\nabla$,
  is a mere abbreviation for $\rD/\rD\rvect$,
  i.e.\ 
  the gradient operator in the $\rvect$-space.
Evidently $\{ \nabla\xi^i \}$ is the dual basis of $\{ \B{i} \}$:
\begin{equation}
  (\B{i}) \cdot \nabla{\xi^j} = {\delta_i}^j    \label{dual}
\end{equation}
  due to the chain rule.
Also
\begin{equation}
  (\nabla{\xi^i}) \otimes \B{i} = \openone    \label{complete}
\end{equation}
  where $\openone$ denotes the unit tensor in the $\rvect$-space.
Note that Eq.~(\ref{complete}) holds thanks to the fact
  that the embedding $\rvect$-space has the same dimension
  as the $\BoldXi$-space (otherwise
  $(\nabla{\xi^i}) \otimes \B{i}$ would be a projection operator
  whose rank is lower than the dimension of the $\rvect$-space).
The dual basis allows us to find the contravariant components
  of a given vector field, say $\mb{f}$, by
\begin{equation}
  f^i = \mb{f}\cdot\nabla{\xi^i}  \label{get-comp-contravariant};
\end{equation}
  substitution of this expression
  into the right-hand side of Eq.~(\ref{contravariant})
  recovers $\mb{f}$ due to Eq.~(\ref{complete}).

As opposed to the contravariant components $(f^i)$
  of a vector field $\mb{f}$,
  we define its covariant components $(f_i)$ by
\begin{equation}
  \mb{f} = f_i \nabla{\xi^i}.
\end{equation}
It is easily confirmed that
\[
  f_i = (\B{i})\cdot\mb{f} = g_{ij} f^j 
\]
  where $g_{ij}$ stands for the Euclidean metric tensor
  defined in Eq.~(\ref{k5}).

Using Eqs.~(\ref{k5}) and (\ref{complete}),
  we identify $(g_{ij})$
  with covariant components of the Euclidean unit tensor,
\[
  g_{ij} 
  \left( \nabla{\xi^i} \right) \otimes \left( \nabla{\xi^j} \right)
  = \openone.
\]
The contravariant components of $\openone$
  comprise the inverse matrix of $(g_{ij})$, denoted by $(g^{ij}$),
  which leads to Eq.~(\ref{k7}).
This implies 
  that $(\tilde{p} g^{ij})$ in Eq.~(\ref{d6})
  and $(K g^{ij})$ in Eq.~(\ref{d7})
  stand for isotropic tensors.

\subsection*{Covariant derivative}

The momentum equation~(\ref{d1}),
  represented in terms of contravariant components,
  contains $\nabla_j$ 
  which generally differs from $\Dj = \rD/\rD{\xi^j}$.
This ``nabla with a subscript''
  is referred to as the \emph{covariant derivative}.
When the space is curved,
  it is not a trivial problem
  to define the covariant derivative
  in an appropriate way.
Fortunately, since the space itself is now flat,
  we can now define $\nabla_j$ as a component of a simple ``gradient''
  using $\nabla = \rD/\rD\rvect$.
For a scalar field, say $\varphi$,
  its gradient is 
\[
  \nabla\varphi = \dd{\varphi}{\rvect}
  = \dd{\xi^i}{\rvect}\,\dd{\varphi}{\xi^i}
  = (\nabla{\xi^i})\,\Di\varphi;
\]
  the covariant derivative of $\varphi$ is given
  by the covariant components (i.e.\ 
  the coefficients for $\nabla\xi^i$) of $\nabla\varphi$,
\begin{equation}
  \nabla_i \varphi = \Di\varphi.
\end{equation}
The covariant derivative of a vector field
  is slightly more complicated.
For $\mb{f}$ given in terms of its contravariant components $(f^i)$,
  the gradient is 
\begin{align*}
  \grad\mb{f}
  &= \nabla\otimes\mb{f} 
  = \left( (\nabla\xi^j)\,\Dj \right) \otimes (f^i\B{i})  \notag \\ 
  &= (\nabla\xi^j) \otimes \Dj (f^i \B{i});
\end{align*}
  we define the covariant derivative $\nabla_j{f^i}$
  by
\begin{equation}
  \Dj (f^i \B{i}) = (\nabla_j {f^i}) \,\B{i} 
  \label{covariant-derivative}
\end{equation}
  so that
\[
  \grad\mb{f} 
  = (\nabla_j {f^i}) \left( (\nabla\xi^j) \otimes \B{i} \right).
\]

A handy way to evaluate $\nabla_j{f^i}$, in the present case,
  is to calculate the Cartesian components of $\mb{f} = f^i\B{i}$
  and then to differentiate them with $\xi^j$,
  which yields the left-hand side of Eq.~(\ref{covariant-derivative}).
For those who disdain to depend on the embedding $\rvect$-space,
  there is a more orthodox way based on a formula
\[
  \nabla_i f^j = \Di f^j + \Gamma_{ik}^j  f^k ,
\]
  with $\Gamma_{ik}^j$ referred to as 
  \emph{Levi-Civita connection} (also known as \emph{Christoffel symbol}
  when it is calculated from $\Di\Dj\rvect$).
Both ways lead to the same result.

The momentum equation~(\ref{d1})
  contains a term arising from the divergence of stress tensor,
\[
  \Div\TensorP 
  = \lim \frac{1}{\Delta{V}}
  \int_{\rD(\Delta{V})}\TensorP\cdot\mb{n}\D{S}
  = \nabla\cdot{\Tp\TensorP}
\]
  where $\Tp(\;\cdot\;)$ denotes
  transposition (practically it could be omitted,
  as $\TensorP$ is symmetric).
Substitution of 
  $\TensorP = P^{ij} (\B{i})\otimes(\B{j})$
  and $\nabla = (\nabla\xi^k)\Dk$
  yields
\begin{align}
 \Div\TensorP 
 &= \dd{}{\rvect} \cdot
    \Tp{\left( P^{ij} (\B{i})\otimes(\B{j}) \right)}  \notag\\ 
 &= (\nabla{\xi^k}) \cdot 
    \Tp{\left\{
      \Dk \left( P^{ij} (\B{i})\otimes(\B{j}) \right)
    \right\}}  \notag \\
 &= \left\{
      \Dk \left( P^{ij} (\B{i})\otimes(\B{j}) \right)
    \right\}
    \cdot \nabla{\xi^k} 
\end{align}
  where the last equal sign
  follows from the definition of the transposition.
At this stage,
  we need the covariant derivative for $(P^{ij})$.
Taking into account a general postulation
  that any formula for a second-order tensor
  should apply to the tensor product of two vectors as well,
  we find the appropriate definition to be
\begin{equation}
  \Dk \left( P^{ij} (\B{i})\otimes(\B{j}) \right)
  = (\nabla_k {P^{ij}})\,(\B{i})\otimes(\B{j})
\end{equation}
  so that
\begin{align*}
 \Div\TensorP  
 &= (\nabla_k {P^{ij}})
 \left( (\B{i})\otimes(\B{j}) \right) \cdot(\nabla{\xi^k}) \notag\\
 &= (\nabla_k {P^{ij}}) (\B{i}) {\delta_j}^k   \notag\\
 &= (\nabla_j {P^{ij}})\,\B{i}.
\end{align*}
Again, $\nabla_k {P^{ij}}$ can be evaluated
  either in terms of the Cartesian components of $\TensorP$
  or with a formula 
\[
  \nabla_k {P^{ij}} 
  = \Dk {P^{ij}} + \Gamma_{kl}^i {P^{lj}} + \Gamma_{kl}^j {P^{il}}.
\]

\subsection*{Velocity and acceleration}

Up to the present point in this appendix,
  we have treated the spatial aspect
  of the mapping from $(\BoldXi,t)$ to $\rvect$ with $t$ fixed.
Now we will detail the temporal aspect
  of this mapping.
Let us recall
  that $\dt$ stands for the Lagrange derivative,
\[
  \dt(\;\cdot\;) = \left( \dd{\;\cdot\;}{t} \right)_{\BoldXi},
\]
  unless specified otherwise (in Eq.~(\ref{*L3}), for example).
The velocity $\mb{v}$
  is then given by Eq.~(\ref{k2a}),
  and the (material) acceleration is
\begin{equation}
  \dt^2 \rvect = \dt\mb{v} = \dt \left( v^i \B{i} \right)
\end{equation}
  as is seen on the left-hand side of the momentum equation
  just above Eq.~(\ref{d1}).
Taking the time-dependence of $\B{i}$ into account,
  we evaluate the acceleration as
\begin{align*}
  \dt\left( v^i \B{i} \right)
  &= (\dt{v^i})\,\B{i} + v^i \dt\B{i}     \notag\\
  &= (\dt{v^i})\,\B{i} + v^i \Di\dt\rvect \notag\\
  &= (\dt{v^i})\,\B{i} + v^i \Di\mb{v}  
\end{align*}
  and rewrite the last term, which contains $\Di\mb{v}$,
  with the covariant derivative.
Thus we find
\begin{equation}
  \dt\mb{v} = \left( \dt v^i + {v^j \nabla_j} v^i \right) \B{i}
  \label{acceleration}.
\end{equation}
The contravariant component of Eq.~(\ref{acceleration}),
  multiplied by $\rho$,
  gives the left-hand side of Eq.~(\ref{d1}).

\subsection*{Derivation of Eq.~(\ref{m1})}

Next, we study a concrete example
  to see how the momentum equation~(\ref{d1}) is evaluated.
With the mapping $\BoldXi \mapsto \rvect$ specified as Eq.~(\ref{u1}),
  the momentum equation~(\ref{d1}) is to be reduced to Eq.~(\ref{m1}).

Eq.~(\ref{u1}) readily yields
  the velocity in Eq.~(\ref{u2}) and the local basis
\begin{equation}
 \Dx\rvect 
    = \begin{bmatrix} 1  \\ 0 \end{bmatrix}_{\Cartesian}, \qquad
 \Dz\rvect 
    = \begin{bmatrix} X' \\ 1 \end{bmatrix}_{\Cartesian}
    \label{u1.basis},
\end{equation}
  where $U$ and $X'$ are understood as
\[
 U  = \dt{X(\ze,t)}, \qquad 
 X' = \Dz{X(\ze,t)}.
\]
Substituting Eq.~(\ref{u1.basis}) into Eq.~(\ref{k5})
  yields $(g_{ij})$ in Eq.~(\ref{u3}).

\begin{widetext}
The natural metric tensor is parametrized as Eq.~(\ref{g1});
  this expression becomes identical to that for $\mathsf{g}$ 
  if $\alpha = 0$ and $\beta = X'$.
The components of the inverse natural metric tensor
  are then
\begin{equation}
 \mathsf{g}_\Nat
  = 
  \begin{bmatrix}
   g_\Nat^{\xi\xi} & g_\Nat^{\xi\ze} \\
   g_\Nat^{\ze\xi} & g_\Nat^{\ze\ze}   
  \end{bmatrix}
  =
  \begin{bmatrix}
   (1+\beta^2) e^\alpha  &  -\beta      \\
   -\beta                &  e^{-\alpha}
  \end{bmatrix}
  \label{g1.inv}.
\end{equation}
By using Eqs.~(\ref{u1.basis}) and (\ref{g1.inv}),  
  the term $g_\Nat^{ij} (\B{i})\otimes(\B{j})$
  in Eq.~(\ref{e6})
  is calculated to be
\begin{align}
 g_\Nat^{ij} (\B{i})\otimes(\B{j})
 &= (1+\beta^2) e^\alpha (\B\xi)\otimes(\B\xi)  
 - \beta \left(
 (\B\xi)\otimes(\B\ze) + (\B\ze)\otimes(\B\xi) 
 \right)                                        
 + e^{-\alpha}  (\B\ze)\otimes(\B\ze)            \notag\\
 &= (1+\beta^2) e^\alpha
 \begin{bmatrix} 
  1 & 0 \\ 0 & 0 
 \end{bmatrix} _{\Cartesian} 
 - \beta 
 \begin{bmatrix} 
  2X' & 1 \\ 1 & 0 
 \end{bmatrix} _{\Cartesian} 
 + e^{-\alpha}
 \begin{bmatrix} 
  {X'}^2 & X' \\ X' & 1 
 \end{bmatrix} _{\Cartesian} \notag\\
 &= 
 \begin{bmatrix} 
  (1+\beta^2) e^\alpha - 2{\beta}X' + e^{-\alpha} {X'}^2 &
  e^{-\alpha} X' - \beta  \\
  e^{-\alpha} X' - \beta  &  e^{-\alpha}
 \end{bmatrix}_{\Cartesian}
 \label{gN.Cartesian}. 
\end{align}
\end{widetext}
Taking notice of the $(x,z)$-component of this expression,
  which corresponds to $\sigma_{xz}/S$,
  we introduce $\tilde\sigma$ given by Eq.~(\ref{sigma}).
Then Eq.~(\ref{e6}) yields
  a concrete expression for $\TensorSigma$ shown in Eq.~(\ref{*P}).

In the present setup,
  the nabla operator is given by
\begin{equation}
 \nabla 
  = \dd{}{\rvect} 
  = \dd{\xi}{\rvect} \Dx + \dd{\ze}{\rvect} \Dz 
\end{equation}
  where
\[
 \dd{\xi}{\rvect} = 
  \begin{bmatrix} 1 \\  {-X'} 
  \end{bmatrix}_{\Cartesian},  \qquad
 \dd{\ze}{\rvect} = 
  \begin{bmatrix} 0 \\  1 
  \end{bmatrix}_{\Cartesian}.
\]
Then the divergence in the momentum equation~(\ref{d1})
  is evaluated 
  in terms of the Cartesian components in the $\rvect$-space:
\begin{align*}
 -\Div\TensorP
 &= 
 -\begin{bmatrix} 1 \\ {-X'} 
  \end{bmatrix}_{\Cartesian} \Dx{\tilde{p}}
 -\begin{bmatrix} 0 \\  1 
  \end{bmatrix}_{\Cartesian} \Dz{\tilde{p}}   \notag\\ &\qquad
 + S
 \left( \Dz 
 \begin{bmatrix} 
  e^\alpha (1 + {\tilde\sigma}^2 ) - 1  &   \tilde\sigma   \\
  \tilde\sigma                          &  e^{-\alpha} - 1 
 \end{bmatrix}_{\Cartesian}
 \right)
 \begin{bmatrix} 0 \\  1 
 \end{bmatrix}_{\Cartesian} 
 \notag\\
 &= 
 -\left(
 \begin{bmatrix} 1 & 0 \\  {-X'} & 1 \end{bmatrix} 
 \begin{bmatrix} \Dx{\tilde{p}} \\ \Dz{\tilde{p}} \end{bmatrix}
 \right)_{\Cartesian}      
 + 
 S \begin{bmatrix}  
    \Dz{\tilde\sigma}  \\  - e^{-\alpha} \Dz\alpha
   \end{bmatrix}_{\Cartesian}
\end{align*}
  where it is taken into account 
  that $\alpha$, $\beta$ and $X'$ are independent of $\xi$.
As for the left-hand side of the momentum equation,
  it is easily shown that
\[
  \dt\mb{v}
  = \begin{bmatrix} {\dt{U}} \\ 0 \end{bmatrix}_{\Cartesian}.
\]

Calculating the inner product of the momentum equation with $\B\xi$, 
  we obtain
\begin{equation}
 \rho\dt{U} 
 = -\Dx{\tilde{p}} + S\Dz{\tilde\sigma} + {\rho} G\sin\theta ;
 \label{EqM.x}
\end{equation}
  similarly, the inner product with $\B\ze$ yields
\begin{align}
 \rho X' \dt{U}
 &= -\Dz{\tilde{p}}
  + S \left( X'\Dz{\tilde\sigma} - e^{-\alpha} \Dz\alpha \right)
 \notag\\ & \quad 
  + \rho G ( X'\sin\theta - \cos\theta ) 
  \label{EqM.z}.
\end{align}
From Eq.~(\ref{EqM.z}), we find
  that $\Dz{\tilde{p}}$ is independent of $\xi$.

Here we use a concrete formulation
  of the free-surface boundary condition 
  for $\TensorP$ (neglecting surface tension
  and surface contamination),
\begin{equation}
 \Ev{P^{ij} n_j}{\ze=H} = \patm g^{ij} n_j   \label{BC.TensorP}
\end{equation}
  where $\patm$ denotes the (constant) atmospheric pressure,
  which can be set equal to zero without loss of generality,
  and $n_j$ stands for the covariant component
  of the surface normal vector,
  which is given by $\mb{n} = \nabla (z-H)$
  so that $n_j = \nabla_j{z} = \rD{z}/\rD{\xi^j}$ 
  for the present case.
For $\tilde{p}$,
  the boundary condition~(\ref{BC.TensorP}) reads
\begin{equation}
  \Ev{(\tilde{p} - \sigma_{zz})}{\ze=H}
   = \patm \,(= 0)                       \label{BC.p}
\end{equation}
  with $\sigma_{zz} = S (e^{-\alpha} - 1)$ 
  according to Eq.~(\ref{*P}).
Evidently,
  Eq.~(\ref{BC.p}) is also independent of $\xi$.
Then $\tilde{p}$ turns out to be totally independent of $\xi$,
  which implies
  that $\Dx{\tilde{p}}$ in Eq.~(\ref{EqM.x}) vanishes,
  leading to Eq.~(\ref{m1}).

\subsection*{Derivation of Eqs.~(\ref{g2}) and (\ref{g3})}

The relaxation of $\mathsf{g}^\Nat$
  is described by Eq.~(\ref{r1}) or, equivalently, Eq.~(\ref{d7}).
We substitute $\mathsf{g}$ parametrized as Eq.~(\ref{u3})
  and $\mathsf{g}^\Nat$ as Eq.~(\ref{g1}) 
  into Eq.~(\ref{d7}),
  together with 
\begin{multline}
 \tau\dt 
 \begin{bmatrix}
  g_\Nat^{\xi\xi} & g_\Nat^{\xi\ze} \\
  g_\Nat^{\ze\xi} & g_\Nat^{\ze\ze}   
 \end{bmatrix}
 \\
 =
 \begin{bmatrix}
  (1+\beta^2) e^\alpha  &  0            \\
  0                     &  -e^{-\alpha}
 \end{bmatrix}
 \tau\dt\alpha
 +
 \begin{bmatrix}
  2\beta e^\alpha  &  -1   \\
  -1               &   0
 \end{bmatrix}
 \tau\dt\beta 
 \notag .
\end{multline}
Equating each component of the matrix
  yields \emph{three} equations 
  for two variables $\alpha$ and $\beta$;
  the equations are consistent (solvable)
  only when $K$ is set appropriately,
  which is calculated, according to Eq.~(\ref{r3}), as
\begin{equation}
 K = \frac{2}{2\cosh{\alpha} + {e^\alpha} {\tilde\sigma}^2 }
  = \frac{2}{2+\Eps}
\end{equation}
  with $\Eps$ given by Eq.~(\ref{Eps}) in the two-dimensional case.
From the $\ze\ze$-component
  and the $\ze\xi$-component of Eq.~(\ref{d7})
  we obtain Eq.~(\ref{g2}) and Eq.~(\ref{g3}), respectively.
\vfill

\section{Variation of the elastic energy $E$}
\label{app:var-E}

Eq.~(\ref{d6}) is obtained 
  from elastic energy $E$ in Eq.~(\ref{e5})
  by calculating its variation
  in regard to $\rvect = \rvect(\BoldXi)$ 
  under the constraint $\detG = 1$.
In this calculation we use
\begin{gather*}
  \delta(\detG) = (\detG) g^{ij} \delta{g_{ij}} , 
  \\
  \delta{g_{ij}} = \delta\left( \B{i} \cdot \B{j} \right)
  = (\Di{\delta\rvect}) \cdot \B{j}
  + \B{i} \cdot (\Dj{\delta\rvect}) ,
  \\
  \intertext{and}
  \\
  \Dj\sqrt{\detG} 
  = {\frac12} \sqrt{\detG} \; g^{kl} \Dj {g_{kl}}
  = \sqrt{\detG} \, (\nabla{\xi^k}) \cdot \Dj\Dk \rvect .
\end{gather*}

\begin{widetext}
The result is as follows:
\begin{align}
  \delta \int E \,\D{V}
  &= {\frac12}S\; \delta 
  \int \left( g_{ij} g_\Nat^{ij} - \nd \right)
  \sqrt{\detG} \,\D^\nd\BoldXi         \notag\\
  &= {\frac12}S  
  \int \left( g_\Nat^{ij} + {\frac12} \Eps g^{ij} \right)
  \delta{g}_{ij}\, 
  \sqrt{\detG} \,\D^\nd\BoldXi         \notag\\
  &= S \int 
  \left( g_\Nat^{ij} + {\frac12}\Eps g^{ij} \right) 
  (\B{i})\cdot(\Dj\delta\rvect)
  \sqrt{\detG} \,\D^\nd\BoldXi         \notag\\
  &= -S \int 
  \left\{ \Dj \left( 
                  \left( g_\Nat^{ij} + {\frac12}\Eps g^{ij} \right) 
                  \sqrt{\detG} \, \B{i}
              \right)
  \right\}
  \cdot \delta\rvect \,\D^\nd\BoldXi         \notag\\
  &= -S \int 
  \left\{ 
     \left( \nabla{\xi^k} \right) \cdot
     \Dk \left( 
            \left( g_\Nat^{ij} + {\frac12}\Eps g^{ij} \right) 
            (\B{i}) \otimes (\B{j})
         \right)
  \right\}  
  \cdot \delta\rvect\,\sqrt{\detG} \,\D^\nd\BoldXi
\end{align}
  and
\begin{align}
  \delta \int p'\,\left( \sqrt{\detG} - 1 \right) \D{V}
  &= \delta 
  \int p'\,\left( \sqrt{\detG} - 1 \right) 
 \sqrt{\detG} \,\D^\nd\BoldXi \notag\\
  &= 
  \int p'\, \,\left( 1 - {\frac{1}{2\sqrt{\detG}}} \right)
  \delta(\det{\mathsf{g}}) \,\D^\nd\BoldXi
  \notag \\
  &= 
  \int p'\, \,\left( \sqrt{\detG} - \frac12 \right)
  2 g^{ij} (\B{i}) \cdot (\Dj{\delta\rvect}) 
  \,\sqrt{\detG} \,\D^\nd\BoldXi     \notag \\
  &= - \int 
  \left\{ \Dj \left(
                 p'\, \left( 2\sqrt{\detG} - 1 \right) g^{ij}
                 (\B{i}) \sqrt{\detG}
              \right)
  \right\} \cdot \delta\rvect \,\D^\nd\BoldXi     \notag \\
  &= - \int 
  \left\{
     \left( \nabla{\xi^k} \right) \cdot
     \Dk \left( 
            p'\, \left( 2\sqrt{\detG} - 1 \right) g^{ij} 
            (\B{i}) \otimes (\B{j})
         \right)
  \right\} \cdot \delta\rvect \,\sqrt{\detG}\,\D^\nd\BoldXi ,
\end{align}
which is summarized as 
\begin{equation}
 \delta \int \left( 
 E -  p'\,\left( \sqrt{\detG} - 1 \right) 
 \right) \D{V}
 = \int 
 \left\{
 (\nabla{\xi^k})\cdot 
 \Dk \left( P^{ij} (\B{i}) \otimes (\B{j}) \right)
 \right\} \cdot \delta\rvect \,\sqrt{\detG}\,\D^\nd\BoldXi 
\end{equation}
  with
\begin{equation}
 P^{ij} = 
 -S \left( g_\Nat^{ij} + {\frac12}\Eps g^{ij}  \right)
 + p'\, \left( 2\sqrt{\detG} - 1 \right) g^{ij} 
 = -S g_\Nat^{ij} + \left( p' - E \right) g^{ij} 
\end{equation}
  where the last equal sign is due to $\sqrt{\detG} = 1$.
Then, rewriting the undetermined multiplier as 
  $p' = \tilde{p} + E - S $, 
  we obtain Eq.~(\ref{d6}).  
\end{widetext}



\begin{thebibliography}{48}
\expandafter\ifx\csname natexlab\endcsname\relax\def\natexlab#1{#1}\fi
\expandafter\ifx\csname bibnamefont\endcsname\relax
  \def\bibnamefont#1{#1}\fi
\expandafter\ifx\csname bibfnamefont\endcsname\relax
  \def\bibfnamefont#1{#1}\fi
\expandafter\ifx\csname citenamefont\endcsname\relax
  \def\citenamefont#1{#1}\fi
\expandafter\ifx\csname url\endcsname\relax
  \def\url#1{\texttt{#1}}\fi
\expandafter\ifx\csname urlprefix\endcsname\relax\def\urlprefix{URL }\fi
\providecommand{\bibinfo}[2]{#2}
\providecommand{\eprint}[2][]{\url{#2}}

\bibitem[{\citenamefont{Miguel and Rubi}(2006)}]{Carmen-Miguel.Book2006}
\bibinfo{author}{\bibfnamefont{M.~C.} \bibnamefont{Miguel}} \bibnamefont{and}
  \bibinfo{author}{\bibfnamefont{M.}~\bibnamefont{Rubi}},
  \emph{\bibinfo{title}{Jamming, Yielding, and Irreversible Deformation in
  Condensed Matter}} (\bibinfo{publisher}{Springer-Verlag},
  \bibinfo{year}{2006}), ISBN \bibinfo{isbn}{3540300287}.

\bibitem[{\citenamefont{Nakahara and Matsuo}(2003)}]{Nakahara.Bussei81}
\bibinfo{author}{\bibfnamefont{A.}~\bibnamefont{Nakahara}} \bibnamefont{and}
  \bibinfo{author}{\bibfnamefont{Y.}~\bibnamefont{Matsuo}},
  \bibinfo{journal}{{Bussei Kenky\^u (Kyoto)}} \textbf{\bibinfo{volume}{81}},
  \bibinfo{pages}{184} (\bibinfo{year}{2003}), \bibinfo{note}{(in Japanese)}.

\bibitem[{\citenamefont{Nakahara and Matsuo}(2005)}]{Nakahara.JPSJ74}
\bibinfo{author}{\bibfnamefont{A.}~\bibnamefont{Nakahara}} \bibnamefont{and}
  \bibinfo{author}{\bibfnamefont{Y.}~\bibnamefont{Matsuo}},
  \bibinfo{journal}{J. Phys. Soc. Japan} \textbf{\bibinfo{volume}{74}},
  \bibinfo{pages}{1362} (\bibinfo{year}{2005}), \eprint{cond-mat/0501447v2}.

\bibitem[{\citenamefont{Nakahara and
  Matsuo}(2006{\natexlab{a}})}]{Nakahara.JStat2006}
\bibinfo{author}{\bibfnamefont{A.}~\bibnamefont{Nakahara}} \bibnamefont{and}
  \bibinfo{author}{\bibfnamefont{Y.}~\bibnamefont{Matsuo}},
  \bibinfo{journal}{J. Stat. Mech.}  (\bibinfo{year}{2006}{\natexlab{a}}),
  \bibinfo{note}{{P07016}}.

\bibitem[{\citenamefont{Nakahara and
  Matsuo}(2006{\natexlab{b}})}]{Nakahara.PRE74}
\bibinfo{author}{\bibfnamefont{A.}~\bibnamefont{Nakahara}} \bibnamefont{and}
  \bibinfo{author}{\bibfnamefont{Y.}~\bibnamefont{Matsuo}},
  \bibinfo{journal}{Phys. Rev. E} \textbf{\bibinfo{volume}{74}},
  \bibinfo{pages}{045102(R)} (\bibinfo{year}{2006}{\natexlab{b}}).

\bibitem[{\citenamefont{Kawazoe et~al.}()\citenamefont{Kawazoe, Kanemura, and
  {Ooshida Takeshi}}}]{Kawazoe}
\bibinfo{author}{\bibfnamefont{H.}~\bibnamefont{Kawazoe}},
  \bibinfo{author}{\bibfnamefont{K.}~\bibnamefont{Kanemura}}, \bibnamefont{and}
  \bibinfo{author}{\bibnamefont{{Ooshida Takeshi}}}, \bibinfo{note}{(under
  preparation)}.

\bibitem[{\citenamefont{Landau and Lifshitz}(1987)}]{Landau.fluid}
\bibinfo{author}{\bibfnamefont{L.~D.} \bibnamefont{Landau}} \bibnamefont{and}
  \bibinfo{author}{\bibfnamefont{E.~M.} \bibnamefont{Lifshitz}},
  \emph{\bibinfo{title}{Fluid Mechanics}}, vol.~\bibinfo{volume}{6} of
  \emph{\bibinfo{series}{Theoretical Physics}}
  (\bibinfo{publisher}{Butterworth-Heinemann}, \bibinfo{year}{1987}).

\bibitem[{\citenamefont{Schlichting and Gersten}(2000)}]{Schlichting.Book2000}
\bibinfo{author}{\bibfnamefont{H.}~\bibnamefont{Schlichting}} \bibnamefont{and}
  \bibinfo{author}{\bibfnamefont{K.}~\bibnamefont{Gersten}},
  \emph{\bibinfo{title}{Boundary Layer Theory}}
  (\bibinfo{publisher}{Springer-Verlag}, \bibinfo{year}{2000}),
  \bibinfo{edition}{8th} ed., ISBN \bibinfo{isbn}{3-540-66270-7}.

\bibitem[{\citenamefont{Joseph}(1990)}]{Joseph.Book1990}
\bibinfo{author}{\bibfnamefont{D.~D.} \bibnamefont{Joseph}},
  \emph{\bibinfo{title}{Fluid Dynamics of Viscoelastic Liquids}}
  (\bibinfo{publisher}{Springer-Verlag}, \bibinfo{year}{1990}).

\bibitem[{\citenamefont{Miyamoto et~al.}(2002)\citenamefont{Miyamoto, Fukao,
  Yamao, and Sekimoto}}]{Miyamoto.PRL88}
\bibinfo{author}{\bibfnamefont{Y.}~\bibnamefont{Miyamoto}},
  \bibinfo{author}{\bibfnamefont{K.}~\bibnamefont{Fukao}},
  \bibinfo{author}{\bibfnamefont{H.}~\bibnamefont{Yamao}}, \bibnamefont{and}
  \bibinfo{author}{\bibfnamefont{K.}~\bibnamefont{Sekimoto}},
  \bibinfo{journal}{Phys. Rev. Letter} \textbf{\bibinfo{volume}{88}},
  \bibinfo{pages}{225504} (\bibinfo{year}{2002}), \eprint{cond-mat/0111005}.

\bibitem[{\citenamefont{Kruse et~al.}(2004)\citenamefont{Kruse, Joanny,
  J{\"u}licher, Prost, and Sekimoto}}]{Kruse.PRL92}
\bibinfo{author}{\bibfnamefont{K.}~\bibnamefont{Kruse}},
  \bibinfo{author}{\bibfnamefont{J.~F.} \bibnamefont{Joanny}},
  \bibinfo{author}{\bibfnamefont{F.}~\bibnamefont{J{\"u}licher}},
  \bibinfo{author}{\bibfnamefont{J.}~\bibnamefont{Prost}}, \bibnamefont{and}
  \bibinfo{author}{\bibfnamefont{K.}~\bibnamefont{Sekimoto}},
  \bibinfo{journal}{Phys. Rev. Letter} \textbf{\bibinfo{volume}{92}},
  \bibinfo{pages}{078101} (\bibinfo{year}{2004}).

\bibitem[{\citenamefont{Kruse et~al.}(2005)\citenamefont{Kruse, Joanny,
  J{\"u}licher, Prost, and Sekimoto}}]{Kruse.EurPhysJE16}
\bibinfo{author}{\bibfnamefont{K.}~\bibnamefont{Kruse}},
  \bibinfo{author}{\bibfnamefont{J.~F.} \bibnamefont{Joanny}},
  \bibinfo{author}{\bibfnamefont{F.}~\bibnamefont{J{\"u}licher}},
  \bibinfo{author}{\bibfnamefont{J.}~\bibnamefont{Prost}}, \bibnamefont{and}
  \bibinfo{author}{\bibfnamefont{K.}~\bibnamefont{Sekimoto}},
  \bibinfo{journal}{Eur. Phys. J. E} \textbf{\bibinfo{volume}{16}},
  \bibinfo{pages}{5} (\bibinfo{year}{2005}).

\bibitem[{\citenamefont{{Ooshida Takeshi} and Sekimoto}(2005)}]{Ooshida.PRL95}
\bibinfo{author}{\bibnamefont{{Ooshida Takeshi}}} \bibnamefont{and}
  \bibinfo{author}{\bibfnamefont{K.}~\bibnamefont{Sekimoto}},
  \bibinfo{journal}{Phys. Rev. Letter} \textbf{\bibinfo{volume}{95}},
  \bibinfo{pages}{108301} (\bibinfo{year}{2005}).

\bibitem[{\citenamefont{Hill}(1950)}]{Hill.Book1950}
\bibinfo{author}{\bibfnamefont{R.}~\bibnamefont{Hill}},
  \emph{\bibinfo{title}{The Mathematical Theory of Plasticity}}
  (\bibinfo{publisher}{Oxford University Press}, \bibinfo{year}{1950}).

\bibitem[{\citenamefont{Hencky}(1924)}]{Hencky.ZAMM4}
\bibinfo{author}{\bibfnamefont{H.}~\bibnamefont{Hencky}},
  \bibinfo{journal}{Zeits. Ang. Math. Mech.} \textbf{\bibinfo{volume}{4}},
  \bibinfo{pages}{323} (\bibinfo{year}{1924}).

\bibitem[{\citenamefont{Bennett}(2006)}]{Bennett.Book2006}
\bibinfo{author}{\bibfnamefont{A.}~\bibnamefont{Bennett}},
  \emph{\bibinfo{title}{Lagrangian fluid dynamics}}
  (\bibinfo{publisher}{Cambridge University Press}, \bibinfo{year}{2006}), ISBN
  \bibinfo{isbn}{0-521-85310-9}.

\bibitem[{\citenamefont{Marsden and Hughes}(1994)}]{Marsden.Book1994}
\bibinfo{author}{\bibfnamefont{J.~E.} \bibnamefont{Marsden}} \bibnamefont{and}
  \bibinfo{author}{\bibfnamefont{T.~J.} \bibnamefont{Hughes}},
  \emph{\bibinfo{title}{Mathematical Foundations of Elasticity}}
  (\bibinfo{publisher}{Dover Publications}, \bibinfo{year}{1994}), ISBN
  \bibinfo{isbn}{0-486-67865-2}, \bibinfo{note}{publised originally by
  Prentice-Hall, 1983}.

\bibitem[{\citenamefont{Nakahara}(1990)}]{Nakahara.Book1990}
\bibinfo{author}{\bibfnamefont{M.}~\bibnamefont{Nakahara}},
  \emph{\bibinfo{title}{Geometry, Topology, And Physics}}
  (\bibinfo{publisher}{Institute of Physics Publishing}, \bibinfo{year}{1990}),
  ISBN \bibinfo{isbn}{0-85274-095-6}.

\bibitem[{\citenamefont{Lee}(1969)}]{Lee.ASME-JAM36}
\bibinfo{author}{\bibfnamefont{E.~H.} \bibnamefont{Lee}},
  \bibinfo{journal}{ASME J. Appl. Mech.} \textbf{\bibinfo{volume}{36}},
  \bibinfo{pages}{1} (\bibinfo{year}{1969}).

\bibitem[{\citenamefont{Lubarda and Lee}(1981)}]{Lee.ASME-JAM48}
\bibinfo{author}{\bibfnamefont{V.~A.} \bibnamefont{Lubarda}} \bibnamefont{and}
  \bibinfo{author}{\bibfnamefont{E.~H.} \bibnamefont{Lee}},
  \bibinfo{journal}{ASME J. Appl. Mech.} \textbf{\bibinfo{volume}{48}},
  \bibinfo{pages}{35} (\bibinfo{year}{1981}).

\bibitem[{\citenamefont{Bingham}(1922)}]{Bingham.Book1922}
\bibinfo{author}{\bibfnamefont{E.~C.} \bibnamefont{Bingham}},
  \emph{\bibinfo{title}{Fluidity and plasticity}}
  (\bibinfo{publisher}{McGraw-Hill}, \bibinfo{address}{New York},
  \bibinfo{year}{1922}).

\bibitem[{\citenamefont{Mei and Yuhi}(2001)}]{Mei.JFM431}
\bibinfo{author}{\bibfnamefont{C.~C.} \bibnamefont{Mei}} \bibnamefont{and}
  \bibinfo{author}{\bibfnamefont{M.}~\bibnamefont{Yuhi}}, \bibinfo{journal}{J.
  Fluid Mech.} \textbf{\bibinfo{volume}{431}}, \bibinfo{pages}{135}
  (\bibinfo{year}{2001}).

\bibitem[{\citenamefont{Duran}(2000)}]{Duran.Book2000}
\bibinfo{author}{\bibfnamefont{J.}~\bibnamefont{Duran}},
  \emph{\bibinfo{title}{Sands, Powders, and Grains; An Introduction to the
  Physics of Granular Materials}} (\bibinfo{publisher}{Springer-Verlag},
  \bibinfo{address}{New York}, \bibinfo{year}{2000}), ISBN
  \bibinfo{isbn}{0-387-98656-1}, \bibinfo{note}{translated by Axel Reisinger}.

\bibitem[{\citenamefont{Press et~al.}(1988)\citenamefont{Press, Flannery,
  Teukolsky, and Vetterling}}]{NumericalRecipe.Book1988}
\bibinfo{author}{\bibfnamefont{W.~H.} \bibnamefont{Press}},
  \bibinfo{author}{\bibfnamefont{B.~P.} \bibnamefont{Flannery}},
  \bibinfo{author}{\bibfnamefont{S.~A.} \bibnamefont{Teukolsky}},
  \bibnamefont{and} \bibinfo{author}{\bibfnamefont{W.~T.}
  \bibnamefont{Vetterling}}, \emph{\bibinfo{title}{Numerical Recipes in C}}
  (\bibinfo{publisher}{Cambridge University Press}, \bibinfo{year}{1988}).

\bibitem[{\citenamefont{Aranson and Tsimring}(2006)}]{Aranson.RMP78}
\bibinfo{author}{\bibfnamefont{I.~S.} \bibnamefont{Aranson}} \bibnamefont{and}
  \bibinfo{author}{\bibfnamefont{L.~S.} \bibnamefont{Tsimring}},
  \bibinfo{journal}{Reviews of Modern Physics} \textbf{\bibinfo{volume}{78}},
  \bibinfo{pages}{641} (\bibinfo{year}{2006}).

\bibitem[{\citenamefont{Sekimoto}(1993)}]{Sekimoto.JNNFM46}
\bibinfo{author}{\bibfnamefont{K.}~\bibnamefont{Sekimoto}},
  \bibinfo{journal}{J. Non-Newtonian Fluid Mech.}
  \textbf{\bibinfo{volume}{46}}, \bibinfo{pages}{219} (\bibinfo{year}{1993}).

\bibitem[{\citenamefont{Carlson and Batista}(1996)}]{Carlson.PRE53}
\bibinfo{author}{\bibfnamefont{J.~M.} \bibnamefont{Carlson}} \bibnamefont{and}
  \bibinfo{author}{\bibfnamefont{A.~A.} \bibnamefont{Batista}},
  \bibinfo{journal}{Phys. Rev. E} \textbf{\bibinfo{volume}{53}},
  \bibinfo{pages}{4153} (\bibinfo{year}{1996}).

\bibitem[{\citenamefont{Nedderman}(1992)}]{Nedderman.Book1992}
\bibinfo{author}{\bibfnamefont{R.~M.} \bibnamefont{Nedderman}},
  \emph{\bibinfo{title}{Statics and Kinematics of Granular Materials}}
  (\bibinfo{publisher}{Cambridge University Press}, \bibinfo{year}{1992}), ISBN
  \bibinfo{isbn}{978-0521404358}.

\bibitem[{\citenamefont{Kamrin and Bazant}(2007)}]{Kamrin.PRE75}
\bibinfo{author}{\bibfnamefont{K.}~\bibnamefont{Kamrin}} \bibnamefont{and}
  \bibinfo{author}{\bibfnamefont{M.~Z.} \bibnamefont{Bazant}},
  \bibinfo{journal}{Phys. Rev. E} \textbf{\bibinfo{volume}{75}},
  \bibinfo{pages}{041301} (\bibinfo{year}{2007}).

\bibitem[{\citenamefont{Sekimoto}(1991)}]{Sekimoto.JNNFM39}
\bibinfo{author}{\bibfnamefont{K.}~\bibnamefont{Sekimoto}},
  \bibinfo{journal}{J. Non-Newtonian Fluid Mech.}
  \textbf{\bibinfo{volume}{39}}, \bibinfo{pages}{107} (\bibinfo{year}{1991}).

\bibitem[{\citenamefont{Ohzono and Shimomura}(2005)}]{Ohzono.PRE72}
\bibinfo{author}{\bibfnamefont{T.}~\bibnamefont{Ohzono}} \bibnamefont{and}
  \bibinfo{author}{\bibfnamefont{M.}~\bibnamefont{Shimomura}},
  \bibinfo{journal}{Phys. Rev. E} \textbf{\bibinfo{volume}{72}},
  \bibinfo{pages}{025203(R)} (\bibinfo{year}{2005}).

\bibitem[{\citenamefont{Landau and Lifshitz}(1986)}]{Landau.elasticity}
\bibinfo{author}{\bibfnamefont{L.~D.} \bibnamefont{Landau}} \bibnamefont{and}
  \bibinfo{author}{\bibfnamefont{E.~M.} \bibnamefont{Lifshitz}},
  \emph{\bibinfo{title}{Theory of Elasticity}}, vol.~\bibinfo{volume}{7} of
  \emph{\bibinfo{series}{Theoretical Physics}} (\bibinfo{year}{1986}).

\bibitem[{\citenamefont{Chaikin and Lubensky}(1995)}]{Chaikin.Book1995}
\bibinfo{author}{\bibfnamefont{P.~M.} \bibnamefont{Chaikin}} \bibnamefont{and}
  \bibinfo{author}{\bibfnamefont{T.~C.} \bibnamefont{Lubensky}},
  \emph{\bibinfo{title}{Principles of condensed matter physics}}
  (\bibinfo{publisher}{Cambridge University Press}, \bibinfo{year}{1995}).

\bibitem[{\citenamefont{McLean}(1962)}]{McLean.Book1962}
\bibinfo{author}{\bibfnamefont{D.}~\bibnamefont{McLean}},
  \emph{\bibinfo{title}{Mechanical properties of metals}}
  (\bibinfo{publisher}{Wiley}, \bibinfo{year}{1962}).

\bibitem[{\citenamefont{Miyazaki and Reichman}(2002)}]{Miyazaki.PRE66}
\bibinfo{author}{\bibfnamefont{K.}~\bibnamefont{Miyazaki}} \bibnamefont{and}
  \bibinfo{author}{\bibfnamefont{D.~R.} \bibnamefont{Reichman}},
  \bibinfo{journal}{Phys. Rev. E} \textbf{\bibinfo{volume}{66}},
  \bibinfo{pages}{050501(R)} (\bibinfo{year}{2002}).

\bibitem[{\citenamefont{Fuchs and Cates}(2002)}]{Fuchs.PRL89}
\bibinfo{author}{\bibfnamefont{M.}~\bibnamefont{Fuchs}} \bibnamefont{and}
  \bibinfo{author}{\bibfnamefont{M.~E.} \bibnamefont{Cates}},
  \bibinfo{journal}{Phys. Rev. Letter} \textbf{\bibinfo{volume}{89}},
  \bibinfo{pages}{248304} (\bibinfo{year}{2002}).

\bibitem[{\citenamefont{Otsuki and Sasa}(2006)}]{Otsuki.JStat2006}
\bibinfo{author}{\bibfnamefont{M.}~\bibnamefont{Otsuki}} \bibnamefont{and}
  \bibinfo{author}{\bibfnamefont{S.}~\bibnamefont{Sasa}}, \bibinfo{journal}{J.
  Stat. Mech.}  (\bibinfo{year}{2006}), \bibinfo{note}{{L10004}}.

\bibitem[{\citenamefont{Frisch}(1995)}]{Frisch.Book1995}
\bibinfo{author}{\bibfnamefont{U.}~\bibnamefont{Frisch}},
  \emph{\bibinfo{title}{Turbulence: the legacy of A.N. Kolmogorov}}
  (\bibinfo{publisher}{Cambridge University Press}, \bibinfo{year}{1995}), ISBN
  \bibinfo{isbn}{0521457130}.

\bibitem[{\citenamefont{Kraichnan}(1965)}]{Kraichnan.PhF8}
\bibinfo{author}{\bibfnamefont{R.~H.} \bibnamefont{Kraichnan}},
  \bibinfo{journal}{Physics of Fluids} \textbf{\bibinfo{volume}{8}},
  \bibinfo{pages}{575} (\bibinfo{year}{1965}).

\bibitem[{\citenamefont{Kaneda}(1981)}]{Kaneda.JFM107}
\bibinfo{author}{\bibfnamefont{Y.}~\bibnamefont{Kaneda}}, \bibinfo{journal}{J.
  Fluid Mech.} \textbf{\bibinfo{volume}{107}}, \bibinfo{pages}{131}
  (\bibinfo{year}{1981}).

\bibitem[{\citenamefont{Kida and Goto}(1997)}]{Kida.JFM345}
\bibinfo{author}{\bibfnamefont{S.}~\bibnamefont{Kida}} \bibnamefont{and}
  \bibinfo{author}{\bibfnamefont{S.}~\bibnamefont{Goto}}, \bibinfo{journal}{J.
  Fluid Mech.} \textbf{\bibinfo{volume}{345}}, \bibinfo{pages}{307}
  (\bibinfo{year}{1997}).

\bibitem[{\citenamefont{Otsuki}(2005)}]{Otsuki.PRE72}
\bibinfo{author}{\bibfnamefont{M.}~\bibnamefont{Otsuki}},
  \bibinfo{journal}{Phys. Rev. E} \textbf{\bibinfo{volume}{72}},
  \bibinfo{pages}{046115} (\bibinfo{year}{2005}).

\bibitem[{\citenamefont{{V. Ya. Shkadov}}(1967)}]{Shkadov.IANMZhG1}
\bibinfo{author}{\bibnamefont{{V. Ya. Shkadov}}}, \bibinfo{journal}{Izv. Akad.
  Nauk. SSSR, Mekh. Zhid. i Gaza} \textbf{\bibinfo{volume}{1}},
  \bibinfo{pages}{43} (\bibinfo{year}{1967}).

\bibitem[{\citenamefont{Ruyer-Quil and
  Manneville}(1998)}]{Ruyer-Quil.EurPhysJB6}
\bibinfo{author}{\bibfnamefont{C.}~\bibnamefont{Ruyer-Quil}} \bibnamefont{and}
  \bibinfo{author}{\bibfnamefont{P.}~\bibnamefont{Manneville}},
  \bibinfo{journal}{Eur. Phys. J. B} \textbf{\bibinfo{volume}{6}},
  \bibinfo{pages}{277} (\bibinfo{year}{1998}).

\bibitem[{\citenamefont{Chang and Demekhin}(2002)}]{Chang.Book2002}
\bibinfo{author}{\bibfnamefont{H.-C.} \bibnamefont{Chang}} \bibnamefont{and}
  \bibinfo{author}{\bibfnamefont{E.}~\bibnamefont{Demekhin}},
  \emph{\bibinfo{title}{Complex Wave Dynamics on Thin Films}}
  (\bibinfo{publisher}{Elsevier}, \bibinfo{year}{2002}).

\bibitem[{\citenamefont{de~Saint-Venant}(1871)}]{Saint-Venant.Paris1871}
\bibinfo{author}{\bibfnamefont{A.~J.~C.} \bibnamefont{de~Saint-Venant}},
  \bibinfo{journal}{C. R. Acad. Sci. Paris} \textbf{\bibinfo{volume}{73}},
  \bibinfo{pages}{147} (\bibinfo{year}{1871}).

\bibitem[{\citenamefont{Forterre and Pouliquen}(2003)}]{Forterre.JFM486}
\bibinfo{author}{\bibfnamefont{Y.}~\bibnamefont{Forterre}} \bibnamefont{and}
  \bibinfo{author}{\bibfnamefont{O.}~\bibnamefont{Pouliquen}},
  \bibinfo{journal}{J. Fluid Mech.} \textbf{\bibinfo{volume}{486}},
  \bibinfo{pages}{21} (\bibinfo{year}{2003}).

\bibitem[{\citenamefont{Bird et~al.}(1987)\citenamefont{Bird, Armstrong, and
  Hassager}}]{Bird.Book1987}
\bibinfo{author}{\bibfnamefont{R.~B.} \bibnamefont{Bird}},
  \bibinfo{author}{\bibfnamefont{R.~C.} \bibnamefont{Armstrong}},
  \bibnamefont{and} \bibinfo{author}{\bibfnamefont{O.}~\bibnamefont{Hassager}},
  \emph{\bibinfo{title}{Dynamics of polymeric liquids}},
  vol.~\bibinfo{volume}{1} (\bibinfo{publisher}{Wiley}, \bibinfo{year}{1987}),
  \bibinfo{edition}{2nd} ed., ISBN \bibinfo{isbn}{047180245X}.

\end{thebibliography}


\end{document}